\RequirePackage{fix-cm}

\documentclass[final]{svjour3} 

\smartqed 

\usepackage{graphicx}
\usepackage{mathptmx} 
\usepackage[]{amsmath} \usepackage[]{amssymb} \usepackage[]{xcolor}
\usepackage[]{subfigure} \usepackage{upgreek} \usepackage{overpic}
\usepackage[]{multirow} \usepackage[]{url} \usepackage[]{nicefrac}
\usepackage[]{bm} \usepackage{import} \usepackage{lipsum}

\DeclareMathAlphabet{\mathcal}{OMS}{cmsy}{m}{n}
\DeclareMathAlphabet\mathbfcal{OMS}{cmsy}{b}{n}
\DeclareMathOperator*{\minimize}{minimize~}
\DeclareMathOperator*{\maximize}{maximize~}
\DeclareMathOperator*{\subjectto}{subject~to~}

\graphicspath{{imgs/}}

\begin{document}

\title{ Data-driven modeling of the chaotic thermal convection in an
  annular thermosyphon }


\author{Jean-Christophe Loiseau }


\institute{ J.-Ch. Loiseau \at
  Laboratoire DynFluid --- Arts et M\'etiers Institute of Technology \\
  151 boulevard de l'h{\^o}pital, 75013 Paris, France. \\
  \email{loiseau.jc@gmail.com} }

\date{Received: date / Accepted: date}

\maketitle


\begin{abstract}
  Identifying accurate and yet interpretable low-order models from data has gained a renewed interest over the past decade.
  In the present work, we illustrate how the combined use of dimensionality reduction and sparse system identification techniques allows us to obtain an accurate model of the chaotic thermal convection in a two-dimensional annular thermosyphon.
  Taking as guidelines the derivation of the Lorenz system, the chaotic thermal convection dynamics simulated using a high-fidelity computational fluid dynamics solver are first embedded into a low-dimensional space using dynamic mode decomposition.
  After having reviewed the physical properties the reduced-order model should exhibit, the latter is identified using SINDy, an increasingly popular and flexible framework for the identification of nonlinear continuous-time dynamical systems from data.
  The identified model closely resembles the canonical Lorenz system, having the same structure and exhibiting the same physical properties.
  It moreover accurately predicts a bifurcation of the high-dimensional system (corresponding to the onset of steady convection cells) occuring at a much lower Rayleigh number than the one considered in this study.

\end{abstract}


\section{Introduction}\label{sec: introduction}

Fluid flows are characterized by high-dimensional nonlinear dynamics
that gives rise to rich structures.  Despite this apparent
complexity, the dynamics often evolves on a low-dimensional
attractor defined by a few dominant coherent structures that contain
significant energy or are useful for
control~\cite{book:holmes:1996}.  Given this property, one might aim
to derive or identify reduced-order models that reproduce
qualitatively and quantitatively the dynamics of the full system.
Over the past decades, identifying robust, accurate and efficient
reduced-order models has thus become a central challenge in fluid
dynamics and closed-loop flow
control~\cite{amr:fabiane:2014,amr:brunton:2015,amr:sipp:2016,arfm:rowley:2017,aiaa:taira:2017}.
Amidst the numerous flows exhibiting such properties, the
buoyancy-driven flow in a thermosyphon is of particular interest
both from a theoretical and practical point of view.  From the
engineering perspective, natural convection in closed loops can play
an important role in the design of thermal energy systems such as
solar heating systems and nuclear reactors.  Natural convection is
also of crucial importance in numerous geophysical situations such
as the mesoscale convective thunderstorms, land and sea breezes
resulting from a differential heating between landmass and an
adjacent body of water, or Hadley cells in the Earth's atmosphere.
From a theoretical point of view finally, natural convection has
attracted a lot of attention ever since the seminal work of Edward
Lorenz~\cite{jas:lorenz:1963} on the derivation of a low-order model
for the Rayleigh-B\'enard convection.
  
As in~\cite{jas:lorenz:1963}, many traditional model reduction
techniques are analytical.
They rely on prior knowledge of the Navier-Stokes equations and
project them onto the span of an othonormal basis of modes,
resulting in a dynamical system in terms of the coefficients of this
expansion basis.
These modes may come from a classical expansion, such as Fourier
modes or Tchebyshev polynomials, or they may be data-driven as in
the proper orthogonal decomposition
(POD)~\cite{qam:sirovich:1987,amr:berkooz:1993}.
Although such approaches have proven to be quite successful for
linear systems~\cite{arfm:rowley:2017}, they have been applied only
with limited success to obtain low-order approximations of nonlinear
systems, mostly on flow oscillators~\cite{jfm:noack:2003}.
Lately, data-driven approaches have been becoming increasingly
popular.
They encompass a wide variety of different approaches such the
eigenrealization algorithm~\cite{jgcd:juang:1985}, the dynamic mode
decomposition~\cite{jfm:rowley:2009,jfm:schmid:2010,jcd:tu:2014} and
its variants, or NARMAX~\cite{book:billings:2013}.
Advances in machine learning are also greatly expanding our ability
to extract governing dynamics purely from data.
One can cite from instance the genetic programming-based system
identification proposed by Bongard \&
Lipson~\cite{pnas:bongard:2007} and Schmidt \&
Lipson~\cite{science:schmidt:2009} or the more recent framework
based on sparsity-promoting regression techniques, SINDy, proposed
by Brunton, Proctor \& Kutz~\cite{pnas:brunton:2016}.
  
Considering a two-dimensional annular thermosyphon, the aim of the
present work is to illustrate how these recent advances in
dimensionality reduction and sparsity-promoting regression
techniques can be harvested to enable the identification of a
low-order Lorenz-like system able to accurately reproduce
qualitatively and quantitatively the key features of chaotic natural
convection in an annular thermosyphon purely from data.
The present paper is organized as follow. First, the exact flow
configuration considered is presented in \textsection~\ref{sec: flow
configuration} along with a brief overview of its chaotic dynamics.
Then, \textsection~\ref{sec: numerics} briefly introduces the reader
to the dimensionality reduction and system identification techniques
used, namely \emph{dynamic mode
decomposition}~\cite{jfm:rowley:2009,jfm:schmid:2010} and the
\emph{sparse identification of nonlinear
  dynamics}~\cite{pnas:brunton:2016}.
Herein, DMD enables us to obtain a low-dimensional representation of the flow's dynamics while SINDy is used to identify an interpretable low-order model correctly approximating these chaotic dynamics.
Finally, the key results of this study are presented in
\textsection~\ref{sec: results} and \textsection~\ref{sec: discussion} while \textsection~\ref{sec:
conclusion} discusses the new perspectives opened by the present
work.


\section{Flow configuration, dynamics and statistical analysis}\label{sec: flow configuration}

Section \textsection\ref{subsec: flow configuration -- flow configuration} introduces the flow configuration and the non-dimensional numbers characterizing our setup.
An overview of the flow dynamics for the set of parameters considered is presented in \textsection\ref{subsec: flow configuration -- dynamics}.
This section also shows strong qualitative evidences that the chaotic dynamics of the flow are closely related to the Lorenz system. Finally, \textsection\ref{subsec: flow configuration -- statistical analysis} summarizes the results of a statistical analysis of the flow dynamics.

\subsection{Flow configuration and non-dimensional numbers}\label{subsec: flow configuration -- flow configuration}

\begin{figure}
  \centering
  \includegraphics[width=.9\textwidth]{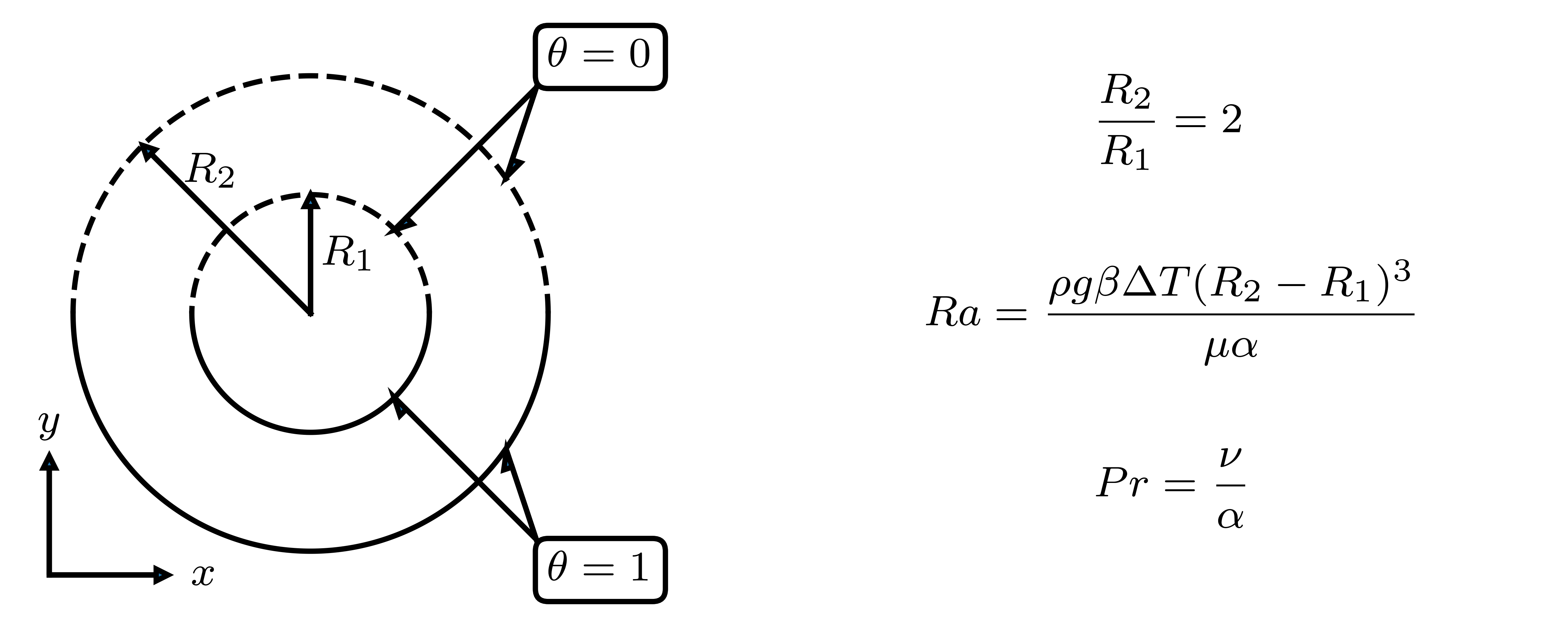}
  \caption{ Left: Sketch of the geometry considered.
    The temperature of the upper walls of the thermosyphon (dashed lines) is set to \(\theta=0\) while it is set to \(\theta=1\) on the lower walls (solid lines).
    The gravity is acting along \(-\bm{e}_y\).
    Right: Definition of the various non-dimensional parameters defining the problem.
    Throughout this manuscript, the Rayleigh number is set to 17 000 while the Prandtl one is set to 5.
  }\label{fig: flow configuration -- geometry}
\end{figure}

The geometry considered consists in two concentric circular enclosures, the inner radius being \( R_1 \) while the outer one is \( R_2 \) as shown in figure~\ref{fig: flow configuration -- geometry}.
In the rest of this work, the ratio of the outer to inner radius is set to
\[
  \frac{R_2}{R_1} = 2.
\]
A constant temperature \( T_0 \) is set at the upper walls while the lower ones are set at a temperature \(T_1 = T_0 + \Delta T\), with \(\Delta T > 0\).
Hereafter, we will work with a non-dimensionalized temperature \(\theta(x, y)\) defined as
\[
  \theta(x, y) = \frac{T(x, y) - T_0}{\Delta T},
\]
where \(x\) and \(y\) are the horizontal and vertical coordinates. Using this non-dimensionalization, the temperature at the lower walls is thus \(\theta_w(y<0)=1\) while the temperature at the upper ones is set to \(\theta_w(y>0) = 0\).
The origin of our reference frame is chosen as the center of the thermosyphon.
Gravity is acting in the vertical direction along \(-\bm{e}_y\) and is characterized by the gravitational acceleration \(g\).

Assuming the working fluid enclosed within the thermosyphon is Newtonian, it is characterized by its density \( \rho \), its dynamic viscosity \( \mu \), its thermal expansion coefficient \(\beta\) and its thermal diffusivity \(\alpha\). Using \(\Delta T\) as the temperature scale and \(R_2 - R_1\) as the length scale, we can then define two non-dimensional parameters, namely the Rayleigh number
\[
  Ra = \frac{\rho g \beta \Delta T \left( R_2 - R_1 \right)^3}{\mu
    \alpha}
\]
and the Prandtl number
\[
  Pr = \frac{\nu}{\alpha},
\]
where \(\nu = \nicefrac{\mu}{\rho}\) is the kinematic viscosity of the working fluid.
Throughout this manuscript, the Rayleigh number is fixed to \(Ra = 17\ 000\) while the Prandtl number is set at \(Pr = 5\).

For the sake of simplicity, we will assume that the flow within the thermosyphon is two-dimensional and incompressible so that the effect of density variations due to temperature can be modeled using the Boussinesq approximation.
Under these assumptions, the dynamics of the flow are governed by the following Navier-Stokes equations
\[
  \begin{aligned}
    \displaystyle \frac{\partial \bm{u}}{\partial t} + \nabla \cdot \left( \bm{u} \otimes \bm{u} \right) & = -\nabla p + \textrm{Pr }\nabla^2 \bm{u} + \textrm{Ra Pr } \theta \bm{e}_y \\
    \displaystyle \frac{\partial \theta}{\partial t} + \left( \bm{u} \cdot \nabla \right) \theta & = \nabla^2 \theta \\
    \displaystyle \nabla \cdot \bm{u} & = 0
  \end{aligned}
\]
where \(\bm{u}(\bm{x}, t)\) is the velocity field, \(p(\bm{x}, t)\) is the pressure field and \(\theta(\bm{x}, t)\) is the temperature one.
Expressing the equations in this form implies that time has been non-dimensionalized with respect to the diffusive temperature time scale.

The mesh consists in 32 spectral elements uniformly distributed in the azimuthal direction and 8 elements uniformly distributed in the radial one.
Within each element, Lagrange interpolants of order 7 based on the Gauss-Lobatto-Legendre quadrature points are used in each direction resulting in 16 384 grid points for the velocity and temperature fields.
Lagrange interpolants of order 5 based on Gauss-Lobatto quadrature points are used for the pressure.
Temporal integration is performed using a third-order accurate scheme and the time-step has been chosen as to satisfy the condition \(\textrm{CFL} \leq 0.5\) during the whole simulation.

\subsection{Flow dynamics}\label{subsec: flow configuration --
  dynamics}

\begin{figure}
  \centering
  \subfigure[Maximum flow rate (clockwise rotation)]{\includegraphics[width=\textwidth]{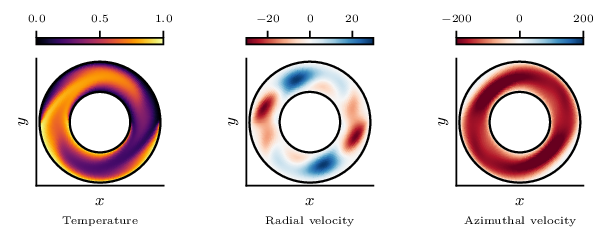}}
  
  \subfigure[Minimum flow rate]{\includegraphics[width=\textwidth]{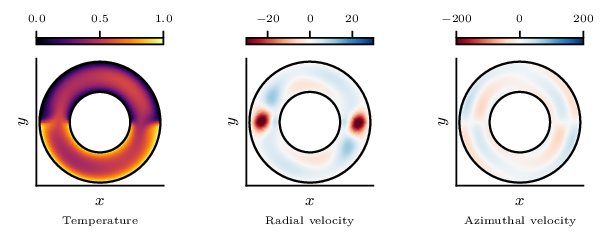}}
  
  \subfigure[Maximum flow rate (counter-clockwise rotation)]{\includegraphics[width=\textwidth]{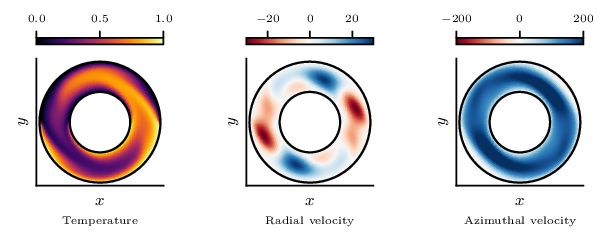}}
  
  \caption{
    Temperature, radial velocity and azimuthal velocity fields at various instants of time.
    The same colorscale has been used for each row.
  }\label{fig: flow configuration -- example flow}
\end{figure}

\begin{figure}
  \centering
  \subfigure[Time-series of the flow rate \(m(t)\)]{\includegraphics[width=\textwidth]{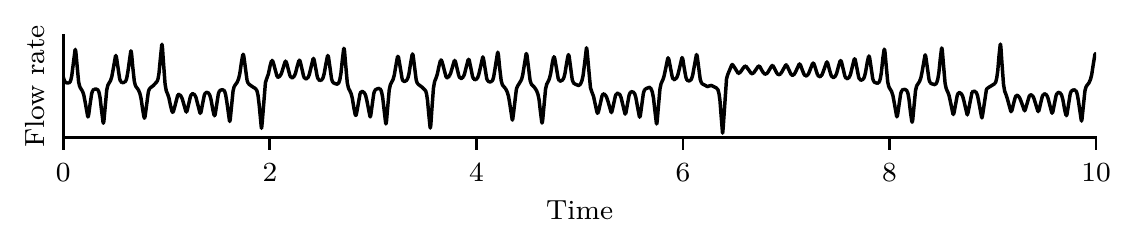}} \\
  \subfigure[Empirical probability distribution of
  \( m(t)
  \)]{\includegraphics[width=.48\textwidth]{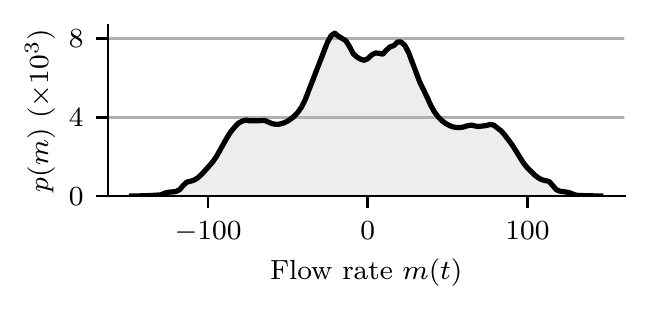}}%
  \hfill
  \subfigure[Power spectral density of \(\dot{m}(t)\)]{\includegraphics[width=.48\textwidth]{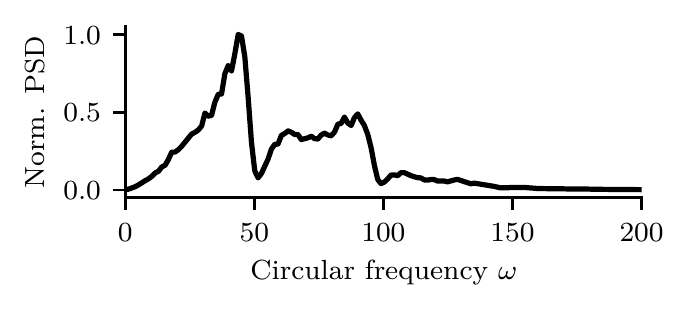}} \\
  \subfigure[Broomhead-King reconstruction of the
  attractor]{\includegraphics[width=\textwidth]{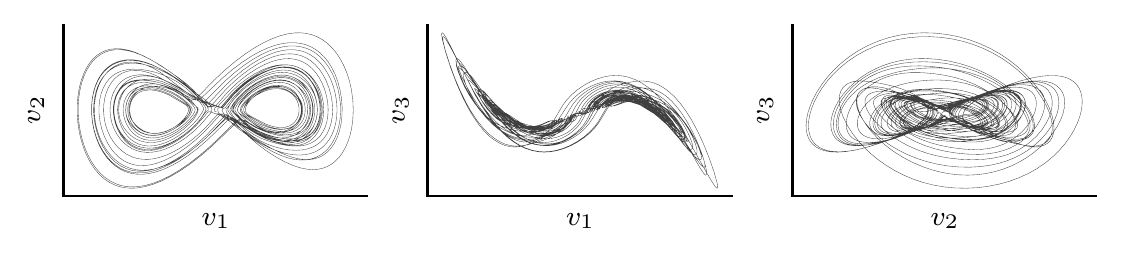}}
  \caption{
    (a) Time-series of the flow rate extracted from a direct numerical simulation of the annular thermosyphon at \( Ra = 17\ 000 \) and \( Pr = 5 \).
    The whole time-series extends from \( t = 0 \) to \( t = 650 \). Only a subset is shown.
    (b) Empirical probability distribution of the flow rate \(m(t)\).
    (c) Power spectral density of the time-derivative \(\dot{m}(t)\) of the flow rate.
    (d) Broomhead-King reconstruction of the underlying strange attractor from the time-series shown in (a).
    The temporal window used for this reconstruction is set to \( \tau = 0.008 \) non-dimensional time-units corresponding to 32 measurements of the flow rate.
    Each panel of figure (b) shows the projection of the strange attractor onto the planes \((v_1, v_2)\), \((v_1, v_3)\) and \((v_2, v_3)\), respectively, where \(v_i\) is the i\textsuperscript{th} Broomhead-King coordinate.
  }\label{fig: flow configuration -- flow rate characterization}
\end{figure}

The Navier-Stokes equations have been integrated forward in time until a statistical steady state is reached.
Figure~\ref{fig: flow configuration -- example flow} depicts the instantaneous temperature (first column), radial velocity (middle column) and azimuthal velocity (right columns) fields at three time instants.
These instantaneous fields highlight that the flow is made of a single convection cell.
This convection cell can either rotate clockwise or counter-clockwise\footnote{A video of these dynamics is available online at \url{https://tinyurl.com/y55buvnc}}.
The dynamics of these seemingly random switches between clockwise and counter-clockwise rotations of the convection cell can be visualized in figure~\ref{fig: flow configuration -- flow rate characterization}(a) depicting a typical time-series of the flow rate \( m(t) \) through a cross-section of the thermosyphon.
Although the convection cell appears to rotate most of the time in the clockwise direction (\(m(t)>0\)) over the time span of figure~\ref{fig: flow configuration -- flow rate characterization}(a), the empirical probability density function of the flow rate computed over more than 650 diffusive time-units and depicted in figure~\ref{fig: flow configuration -- flow rate characterization}(b) shows that both rotations are equally likely.
Figure~\ref{fig: flow configuration -- flow rate characterization}(a) shows moreover that the convection cell appears to oscillate at the same characteristic frequency whether it rotates clockwise or counter-clockwise.
Looking at the power spectral density of the time-derivative \( \dot{m}(t) \) of the flow rate in figure~\ref{fig: flow configuration -- flow rate characterization}(c), this characteristic frequency is estimated to be \( \omega \simeq 45.36 \) corresponding to a period \( \tau = 0.138 \).
Although no linear stability analysis of the Navier-Stokes equations has been conducted during this study, we hypothesize that this characteristic frequency is close to the frequency of the leading eigenmode of the linearly unstable steady clockwise (or counter-clockwise) convection cell.
Confirmation of this hypothesis is left for future studies as it would require the introduction of additional analyses beyond the scope of the present work.

\begin{figure}
  \centering
  \subfigure[ Time-series of \( x(t) \)]{\includegraphics[width=\textwidth]{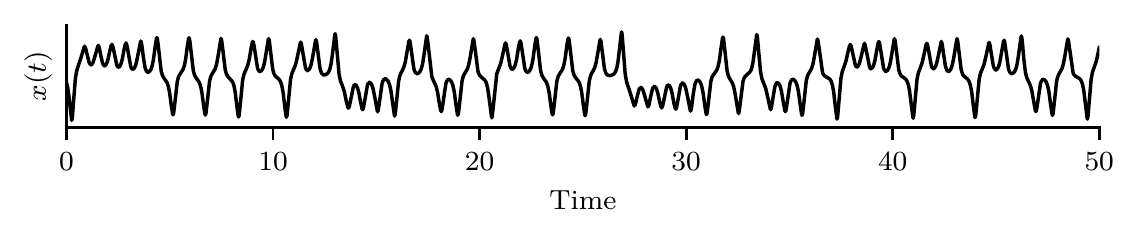}} \\
  \subfigure[ Broomhead-King reconstruction of the
  attractor]{\includegraphics[width=\textwidth]{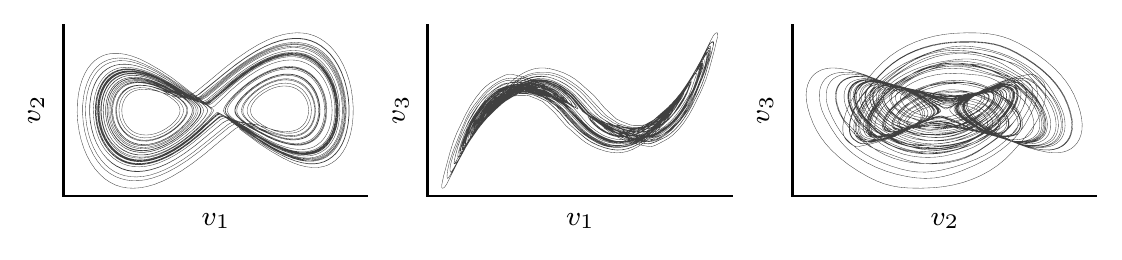}}
  \caption{
    (a) Time-series of the \( x \) variable of the Lorenz system for \((\sigma, \rho, \beta) = (10, 28, \nicefrac{8}{3})\).
    The whole time-series extends from \( t = 0 \) to \( t = 200 \). Only a subset is shown.
    (b) Broomhead-King reconstruction of the underlying strange attractor from the time-series shown in (a).
    The temporal window used for this reconstruction is set to \( \tau = 0.032 \) non-dimensional time-units corresponding to 32 measurements of the \(x\) variable.
    Each panel of figure (b) shows the projection of the strange attractor onto the planes \((v_1, v_2)\), \((v_1, v_3)\) and \((v_2, v_3)\), respectively, where \(v_i\) is the i\textsuperscript{th} Broomhead-King coordinate.
  }\label{fig: flow configuration -- lorenz system}
\end{figure}

A natural question to ask is whether these switches between clockwise and counter-clockwise rotations of the convection cell are governed by a random stochastic process or by an unknown deterministic one.
To answer to this question, one can use tools from dynamical systems theory such as the Takens embedding theorem~\cite{takens:1981} or the Broomhead-King attractor reconstruction technique~\cite{physicaD:broomhead:1986} which combines ideas from the Takens embedding theorem with the singular value decomposition of a Hankel matrix constructed from time-lagged versions of a single time-series.
In this work, the latter has been preferred as it enables the reconstruction of the underlying attractor from a single time-series while also providing a simple diagnostic to determine its dimension by looking at the singular values distribution of the Hankel matrix.
For more details about this attractor reconstruction technique, interested readers are refered to the original paper~\cite{physicaD:broomhead:1986} as well as to Appendix~\ref{appendix: broomhead-king}.
Using the time-series of the flow rate, the attractor reconstructed using the Broomhead-King technique is shown in figure~\ref{fig: flow configuration -- flow rate characterization} (d).
The distribution of singular values (not presented) strongly suggest that this attractor is only three-dimensional.
Additionally, the Poincar\'e section (not shown) highlights that, even though this attractor is embedded within a three-dimensional space, it is almost two-dimensional.
Closer inspection of the Poincar\'e section indicates that it moreover has a fractal structure with a Haussdorf dimension slightly larger than 2, albeit longer simulations would be required to precisely determine this fractal dimension.
These empirical results thus strongly suggest that, despite the high-dimensionality of the discretized Navier-Stokes equations, the dynamics of the convection cell are governed by an underlying low-dimensional deterministic process exhibiting chaos.
The vast majority of this work is thus dedicated to identifiying this low-dimensional chaotic system from data obtained by direct numerical simulations.
It must be noted finally that the attractor reconstructed in figure~\ref{fig: flow configuration -- flow rate characterization}(d) looks strikingly similar to the one obtained when applying the same technique to the famous Lorenz system, see figure~\ref{fig: flow configuration -- lorenz system}.
This striking resemblance let us think that the unknown low-dimensional system we look for may be very similar to the Lorenz system.
This plausible analogy, already noted by various other authors~\cite{physd:yorke:1987,jfm:ehrhard:1990,ijhmt:louisos:2015} when studying similar flows, will guide all of the coming analyses.

\subsection{Statistical analysis}\label{subsec: flow configuration -- statistical analysis}

\begin{figure}
  \centering
  \subfigure[Mean flow]{\includegraphics[width=\textwidth]{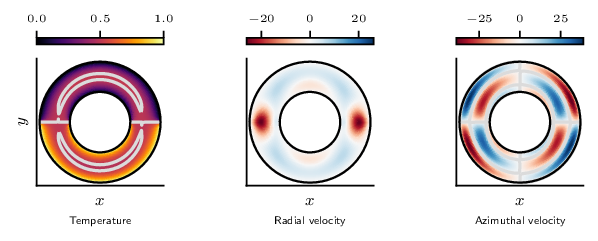}}
  \subfigure[Diagonal components of the Reynolds stress tensor]{\includegraphics[width=\textwidth]{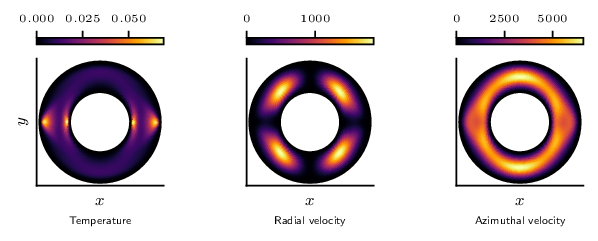}}
  \caption{
    (a) Mean flow computed over 650 non-dimensional time units.
    The light gray line in the temperature plot depicts the \( \theta(x, y) = 0.5 \) isocontour.
    For the azimuthal velocity field, it depicts the \( u_a(x, y) = 0 \) isocontour.
    (b) Diagonal entries of the Reynolds stress tensor corresponding to (d) \( \overline{\theta^{\prime} \theta^{\prime}} \), (e) \( \overline{u_r^{\prime} u_r^{\prime}} \) and (f) \( \overline{u_a^{\prime} u_a^{\prime}} \) where prime variables denote fluctuating quantities.
  }\label{fig: statistical analysis -- mean flow}
\end{figure}

Before diving in the identification of the unknown low-dimensional system governing the dynamics of the convection cell, let us first gain more insights by characterizing more quantitatively the statistics of the system.
Figure~\ref{fig: statistical analysis -- mean flow} shows the mean state and the diagonal
entries of the corresponding Reynolds stress tensor.
These have been obtained by averaging over more than 650 diffusive time-units the various fields once the simulation had reached a statistical steady state.
All of these quantities are either symmetric or anti-symmetric with respect to both the vertical and the horizontal axes.
Given that the fixed point of the Navier-Stokes equations has not been computed, we cannot conclude regarding the similiarties of the base flow and the mean flow albeit we expect these two to be relatively similar.
Looking at the bottom row of figure~\ref{fig: statistical analysis -- mean flow}, it is worthy to note moreover that, although the fluctuations of azimuthal velocity are relatively invariant in the azimuthal direction, the temperature fluctuations are essentially localized in the vicinity of \( y = 0 \), i.e.\ where the temperature applied onto the walls jumps from \( \theta_w = 1 \) to \( \theta_w = 0 \).

\begin{figure}
  \centering \includegraphics[width=\textwidth]{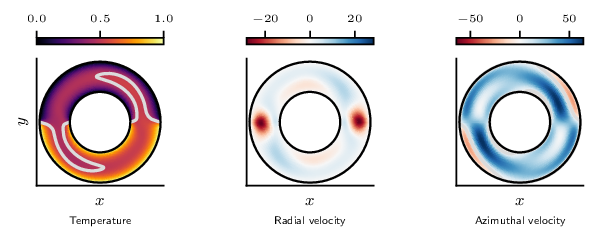}
  \caption{
    Conditionally averaged mean flow \( \bar{\bm{u}}_{CCW}(\bm{x}) \) computed over 650 non-dimensional time units.
    The light gray line in the temperature plot depicts the \( \theta(x, y) = 0.5 \) isocontour.
    For the azimuthal velocity field, it depicts the \( u_a(x, y) = 0 \) isocontour.
  }\label{fig: statistical analysis -- conditional mean flows}
\end{figure}

Let us now look more closely at the organization of the flow when it rotates either in the clockwise or counter-clockwise direction. Figure~\ref{fig: statistical analysis -- conditional mean flows} depicts the the mean flow conditionned on the negative flow rate, i.e.
\[
  \bar{\bm{u}}_{CCW}(\bm{x}) = \mathbb{E} \left[ \bm{u}(\bm{x}, t) \vert m(t) < 0 \right],
\]
As expected, it describes a convection cell rotating in the counter-clockwise direction.
Similarly, \( \bar{\bm{u}}_{CW}(\bm{x}) = \mathbb{E} \left[ \bm{u}(\bm{x}, t) \vert m(t) > 0 \right] \) corresponds to a convection cell rotating in the clockwise direction (not shown).
Even though their amplitudes change, the spatial distribution of the Reynolds stresses associated to this conditionally averaged mean flow appears quite similar that of the true mean flow (not shown).
These two conditional mean flows are moreover symmetric of one another with respect to the vertical axis, consistent with the observation made from the empirical probability density function of the flow rate that the clockwise and counter-clockwise rotations of the convection cell are equally likely.
Within the embedded phase space, these two conditional mean states are located approximately at the center of the empty regions on each side of the attractor (not shown), thus suggesting that they may be very similar to the linearly unstable steady solutions of the Navier-Stokes equations corresponding to the two configurations.
From a dynamical system point of view, we expect these two points to be saddle-foci, i.e.\ linearly unstable fixed points having a single complex conjugate pair of unstable eigenvalues while all the others are stable (i.e.\ negative real part).
Once again, confirmation of this hypothesis is left for future studies.

\begin{figure}
  \centering
  \subfigure[]{\includegraphics[width=\textwidth]{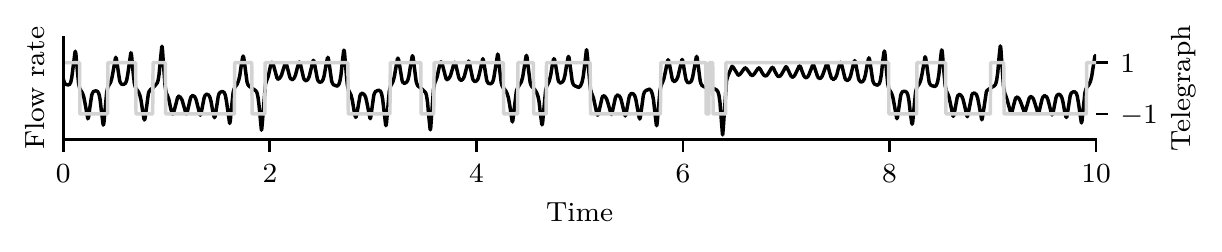}} \\
  \subfigure[]{\includegraphics[width=.48\textwidth]{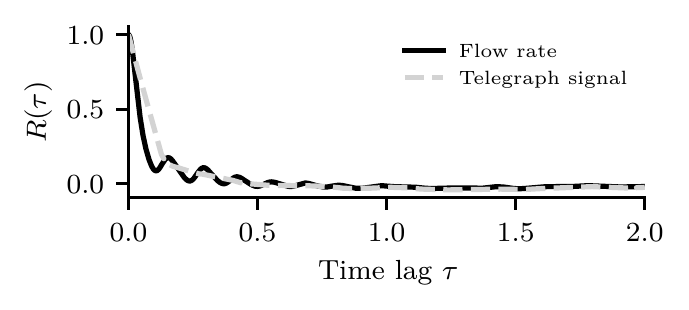}}
  \hfill
  \subfigure[]{\includegraphics[width=.48\textwidth]{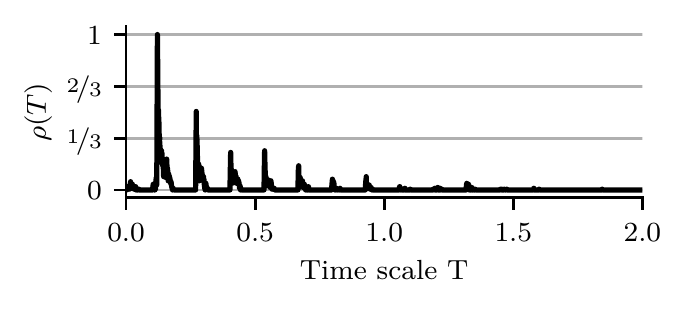}}
  \caption{
    (a) Time-series of the flow rate \(m(t)\) (gray) recorded from direct numerical simulation and the corresponding telegraph signal (gray).
    (b) Autocorrelation function \(R_{mm}(\tau)\) of the flow rate (gray) and \(R_{ss}(\tau)\) of the telegraph singal (gray dash).
    (c) Distribution of the impulse duration of the telegraph signal.
  }\label{fig: statistical analysis -- telegraph signal}
\end{figure}

Finally, let us look back once more at the time-series of the flow rate \( m(t) \) and inspect further its statistical properties.
For that purpose, we will use symbolic dynamics by introducing the telegraph signal \( s(t) \).
It is defined as
\[
  s(t) = \text{sign}\left( m(t) \right),
\]
i.e.\ \( s(t) = 1 \) if the convection cell rotates clockwise and \( s(t) = -1 \) otherwise.
A time-series of this telegraph signal \(s(t)\) along with the corresponding evolution of the flow rate \(m(t)\) is shown in figure~\ref{fig: statistical analysis -- telegraph signal}(a).
By discarding the oscillatory behaviour in the vicinity of the saddle-foci, the introduction of the telegraph signal \( s(t) \) thus enables us to focus our attention onto the switching dynamics only.
Looking at the autocorrelation functions of the signals depicted in figure~\ref{fig: statistical analysis -- telegraph signal}(b), it can be observed that the autocorrelation decay observed at short time (i.e.\ \( \tau < 0.25 \)) is primarily dictated by the switching dynamics.
One can additionally look at the statistical distribution of the duration of the clockwise and counter-clockwise rotations.
This distribution is shown in figure~\ref{fig: statistical analysis -- telegraph signal}(c).
It is close to being a discrete distribution with the peaks being located at constant intervals of time from one another.
The most likely duration of rotation in one direction is \( \tau_0 \simeq 0.12 \), consistent with the frequency measured from the power spectral density of \( \dot{m}(t) \) in figure~\ref{fig: flow configuration -- flow rate characterization}(c) and the characteristic decay time scale of the autocorrelation functions.
This characteristic duration corresponds to the typical time it takes the system to perform a complete oscillation around one of the two sadde-foci.
The other peaks in the distribution in figure~\ref{fig: statistical analysis -- telegraph signal}(c) thus correspond to events during which the convection cell oscillates two or more times before it switches direction.
The amplitude of the peaks at \( \tau_0 \), \( 2\tau_0 \), \( 3\tau_0 \), etc appears to decay exponentially fast thus highlighting that the convection cell is unlikely to oscillate more than a handful of times before switching direction.
All of the statistical properties just described are thus properties that the low-dimensional system we aim to identify would need to verify before we can claim that it captures the essence of the thermosyphon's chaotic dynamics.
It should be noted finally that very similar results have been reported in~\cite{physicaA:anishchenko:2003} when studying the Lorenz system, hence further highlighting the possible connections that exist between the two systems.

\section{From Rayleigh-B\'enard to the Lorenz system}\label{sec: lorenz}

This section essentially is of pedagogical interest.
Given the similarities highlighted in the previous section between our flow configuration and the classical Lorenz system, its aims is to show how one can derive the Lorenz system from the equations governing the Rayleigh-B\'enard convection between two infinite horizontal plates.
Additionally, this derivation will provides us with guidelines for the identification of the unknown low-dimensional model in \textsection\ref{sec: results}.
Readers already familiar with this derivation may skip directly to \textsection\ref{subsec: lorenz -- properties} where the properties of the Lorenz are discussed.

\subsection{Rayleigh-B\'enard convection}

The classical setup for Rayleigh-B\'enard convection is that of a horizontal layer of fluid heated from below and cooled from above in which the fluid develops a regular pattern of convection cells known as B\'enard cells.
From a mathematical point of view, the dynamics of the flow are governed by the same set of equations as the thermosyphon with the exception that the length scale used to define the non-dimensional numbers is now chosen as the height \( H \) between the two plates.
As discussed in the introduction, this classical setup for Rayleigh-B\'enard convection serves as a simplified model in numerous fields (e.g.\ atmospheric convection in meteorology).

Assuming that the lower wall is at temperature \( \theta(y=0) = 1 \) and the upper one \( \theta(y=0) = 0 \), it can easily be shown that the equations admit the following fixed point
\[
  \bm{u}(\bm{x}) = 0 \quad \text{and} \quad \Theta(\bm{x}) = 1 - y.
\]
This solution corresponds to a pure conducting state.

In order to simplify our derivation of the Lorenz system, let us now introduce the streamfunction \( \psi(\bm{x}, t) \) defined such that
\[
  \bm{u}(\bm{x}, t) = \nabla \times \psi(\bm{x}, t) \bm{e}_z
\]
with appropriate boundary conditions.
Rewriting the Navier-Stokes equations in terms of the fluctuations with respect to this base state yields
\begin{equation}
  \begin{aligned}
    & \displaystyle \frac{\partial}{\partial t} \nabla^2 \psi = -\mathcal{J}(\psi, \nabla^2 \psi) + Ra Pr \frac{\partial \theta}{\partial x} + Pr \nabla^4 \psi\\
    & \displaystyle \frac{\partial \theta}{\partial t} =
    -\mathcal{J}(\psi, \theta) + \nabla^2 \theta + \frac{\partial
      \psi}{\partial x}
  \end{aligned}
  \label{eq: lorenz -- rayleigh-benard}
\end{equation}
where the nonlinear terms are expressed as a Jacobian operator given by
\[
  \mathcal{J}(f, g) = \frac{\partial f}{\partial x}\frac{\partial
    g}{\partial y} - \frac{\partial g}{\partial x}\frac{\partial
    f}{\partial y}.
\]
Expressing everything in terms of the fluctuations to the base state enables us to simply impose a zero Dirichlet boundary conditions for the temperature fluctuation \( \theta(\bm{x}) \) on both walls while, as Lorenz, we will impose free-slip boundary conditions for the velocity.
Finally, periodic boundary conditions are assumed in the horizontal direction.

\subsection{Truncated Galerkin expansion and the Lorenz system}

Saltzman~\cite{jas:saltzman:1962} has shown that \( \psi(\bm{x}, t) \) and \( \theta(\bm{x}, t) \) in Eq.~\eqref{eq: lorenz -- rayleigh-benard} can be expressed as infinite Fourier series in both \( x \) and \( y \).
Rather than using these infinite Fourier series which would yield intractable calculations, let us rather consider a three-terms truncated Galerkin expansion given by
\[
  \begin{aligned}
    \psi(x, y, t) & = a(t) \sin(\pi y) \sin(k \pi x) + \cdots \\
    \theta(x, y, t) & = b(t) \sin(\pi y) \cos(k \pi x) + c(t)
    \sin(2\pi y) + \cdots
  \end{aligned}
\]
The scalars \( a(t) \) and \( b(t) \) are the amplitudes of the velocity and temperature fields of the convection cells with wavenumber \( k \) in the horizontal direction while \( c(t) \) describes the modification of the mean temperature profile due to the onset of convection.

Introducing this ansatz into Eq.~\eqref{eq: lorenz -- rayleigh-benard} and performing a Galerkin projection yields
\[
  \begin{aligned}
    \frac{\mathrm{d}a}{\mathrm{d}t} & = -Pr (k^2 + \pi^2) a + Pr Ra \frac{k}{k^2 + \pi^2} b \\
    \frac{\mathrm{d}b}{\mathrm{d}t} & = \pi k ac + k a - (k^2 + \pi^2) b \\
    \frac{\mathrm{d}c}{\mathrm{d}t} & = k \frac{\pi}{2}ab - 4\pi^2 Pr\
    c.
  \end{aligned}
\]
After some change of variables, the above model can be transformed into the canonical Lorenz system
\[
  \begin{aligned}
    \frac{\mathrm{d}x}{\mathrm{d}t} & = \sigma \left( y - x \right) \\
    \frac{\mathrm{d}y}{\mathrm{d}t} & = x \left( \rho - z \right) - y \\
    \frac{\mathrm{d}z}{\mathrm{d}t} & = xy - \beta z
  \end{aligned}
\]
where \( \sigma \) plays the role of the Prandtl number, \( \rho \) that of the Rayleigh number and \( \beta \) characterizes the aspect ratio of the convection cells.
For more details about this derivation, interested readers are referred to \cite{jas:saltzman:1962}.

\subsection{Properties of the Lorenz system}\label{subsec: lorenz -- properties}

A very large number of studies have been dedicated to the Lorenz system over the past 50 years.
The discussion that comes thus does not aim to delve extensively into all of the properties of the Lorenz system, but rather at highlighting the key properties we expect the to-be-identified low-order model to exhibit if the chaotic dynamics of the thermosyphon are indeed Lorenz-like.

The first property of interest is that the Lorenz system belong to the class of dissipative dynamical systems, i.e.\ volumes in phase space contract under the flow \( \bm{f}(x, y, z) \).
Denoting by \(V_0\) a volume of nearby initial conditions, it can easily be shown that its evolution is governed by
\[
  \frac{\mathrm{d}V}{\mathrm{d}t} = -\left( \sigma + 1 + \beta \right)
  V
\]
where \( -\left( \sigma + 1 + \beta \right) = \nabla \cdot \bm{f}(x, y, z) \).
The solution of this ordinary differential equation is
\[
  V(t) = e^{-\left( \sigma + 1 + \beta \right)t} V_0.
\]
Given that \( \sigma, \beta > 0\), volumes in phase space thus shrink exponentially fast to a set of measure zero (i.e.\ fixed points, periodic orbits or strange attractor).
When identifying the low-order model in \textsection\ref{sec: results}, such a condition will impose some constraints on the set of admissible coefficients of the model.

The second property of interest is that the Lorenz equations are equivariant with respect to the transformation
\[
  (x, y, z) \to (-x, -y, z).
\]
This equivariance property implies that all solutions are either symmetric or have a symmetric partner.
From a physical point of view, it describes the fact that the clockwise and counter-clockwise rotations of the convection cell are symmetric from one another and equally likely.
Additionally, as will be discussed in \textsection\ref{sec: results}, this property also limits the pool of functions one can use for system identification.

Finally, the last property of interest is related to the nature of the
z-axis.  Setting
\[x = y = 0 \quad \forall t,\]
the Lorenz system reduces to
\[
  \frac{\mathrm{d}z}{\mathrm{d}t} = -\beta z
\]
and its solution is simply
\[
  z(t) = e^{-\beta t} z_0.
\]
The z-axis thus forms an invariant set and belongs to the stable manifold of the fixed point located at the origin of the phase space.
Physically, this mathematical property describes the fact that, starting from a temperature distribution uniform in the horizontal direction and forcing the fluid to be at rest, the system will naturally relaxes exponentially rapidly to the pure conducting state characterized by a linear temperature profile in the vertical direction and uniform in the horizontal one.
From our point of view, the three aforementioned mathematical properties are the fundamental ones the to-be-identified system needs to verify (along with the statistical properties presented in \textsection\ref{subsec: flow configuration -- statistical analysis}) to be physically consistent.
After this pedagogical interlude, the rest of this work now focuses on the numerical methods and the actual identification of our unknown low-order model.


\section{Numerical methods for low-dimensional system identification}\label{sec: numerics}

In the derivation of his famous equations, Lorenz has used prior knowledge of the governing equations and of their mathematical properties to find a natural system of coordinates (i.e.\ the leading Fourier-Fourier modes) to represent the state of the system and to derive the equations governing its dynamics within this low-dimensional space.
In the present work, the singularity of our temperature boundary condition at \( y = 0 \) prevents us for doing so.
As a consequence, we thus need to turn our attention to dimensionality reduction and system identification techniques to find a proper low-dimensional embedding of the state and approximate/identify its governing equations within this low-dimensional space.
To do so, \textsection\ref{subsec: numerics -- dmd} introduces the reader to the \emph{dynamic mode decomposition} (DMD), an increasingly popular dimensionality reduction proposed by Schmid~\cite{jfm:schmid:2010} in 2010.
Similarly, \textsection\ref{subsec: numerics -- sindy} provides a rapid introduction to the \emph{sparse identification of nonlinear dynamics} (SINDy), a new framework for the identification of continuous-time systems proposed by Brunton \emph{et al.}~\cite{pnas:brunton:2016}.
Using DMD on our data will thus provides us with a low-dimensional representation of the flow's evolution from which a low-order model will be fitted using SINDy.

\subsection{Dynamic Mode Decomposition}\label{subsec: numerics -- dmd}

Dynamic mode decomposition (DMD) is an increasingly popular dimensionality reduction technique that has originated from the field of fluid dynamics~\cite{jfm:rowley:2009,jfm:schmid:2010}.
Since its introduction, numerous variants have been proposed, see for instance~\cite{jfm:rowley:2009,jfm:schmid:2010,jcd:tu:2014,pof:jovanovic:2014,jns:williams:2015,book:kutz:2016,expfluids:dawson:2016,jfm:noack:2016,siam:leclainche:2017,arxiv:hirsh:2019}.
It has also been related to Koopman theory~\cite{pnas:koopman:1932}.

Considering a zero-mean sequence of evenly sampled \( m \)-dimensional snapshots \( \left\{ \bm{q}(\bm{x}, t_k) \right\}_{k=1, n} \), DMD aims at finding a linear operator \( \bm{A} \in \mathbb{R}^{m \times m }\) solution to
\[
  \begin{aligned}
    \minimize_{\bm{A}} \sum_{k=1}^{n-1} \| \bm{q}_{k+1} - \bm{Aq}_k
    \|_2^2
  \end{aligned}
\]
where \( \bm{q}_k = \bm{q}(\bm{x}, t_k) \).
Introducing the data matrix
\[
  \bm{X} = \begin{bmatrix} \bm{q}_1 & \bm{q}_2 & \cdots &
    \bm{q}_{n-1} \end{bmatrix}
\]
and its time-shifted counterpart
\[
  \bm{Y} = \begin{bmatrix} \bm{q}_2 & \bm{q}_3 & \cdots &
    \bm{q}_n \end{bmatrix},
\]
the minimization problem can be recast in matrix form as
\[
  \minimize_{\bm{A}} \| \bm{Y} - \bm{AX} \|_F^2.
\]
Its solution is simply the least-squares solution
\[
  \bm{A}_{LS} = \bm{C}_{yx} \bm{C}_{xx}^{-1},
\]
where \( \bm{C}_{xx} = \bm{XX}^H \) and \( \bm{C}_{yx} = \bm{YX}^H \) are sample variance-covariance matrices.
Unfortunately, in fluid mechanics we are often in the situation where \( m \gg n \), i.e.\ the dimension of the state vector \( \bm{q} \) is generally much larger than the number of samples we are able to collect.
Hence, the estimates \( \bm{C}_{xx} \) and \( \bm{C}_{yx} \) of the variance-covariance matrices are not converged, implying that \( \bm{C}_{xx} \) is likely to be rank-deficient and its inverse ill-posed.
As a consequence, the least-square solution \( \bm{A}_{LS} \) tends to be overly sensitive to minute details of the data used to train the model and to be not statistically representative of the dynamics one tries to describe.

In order to circumvent this issue, it is common practice in regression analysis to regularize the optimization problem by assuming for instance that the unknown matrix \( \bm{A} \) is low-rank.
If so, it can be factorized as \( \bm{A} = \bm{PQ}^H \), with \( \bm{P} \) and \( \bm{Q} \in \mathbb{R}^{m \times r} \) two rank-\(r\) matrices providing bases for the columnspan and the rowspan of \( \bm{A} \), respectively.
Without loss of generality, one can additionally assumes that the columns of \( \bm{P} \) form an orthonormal set of \( m \)-dimensional vectors, i.e.\ \( \bm{P}^H \bm{P} = \bm{I} \).
Under these assumptions, the regularized DMD problem then reads
\[
  \begin{aligned}
    \minimize_{\bm{P}, \bm{Q}} & \| \bm{Y} - \bm{PQ}^H \bm{X} \|_F^2 \\
    \subjectto & \text{rank } \bm{P} = r, \\
    & \text{rank } \bm{Q} = r, \\
    & \bm{P}^H \bm{P} = \bm{I}.
  \end{aligned}
\]
Although the above optimization problem is non-convex due to the rank constraint, it belongs to a wider class of problems known as \emph{Reduced Rank Regression} (RRR) whose properties and closed-form solutions have been studied as early as the mid 1970's by Izenman~\cite{jma:izenman:1975} and recently reviewed by De la Torre~\cite{ieee:delatorre:2012}.
It can be shown that the above norm minimization problem is equivalent to the following trace maximization problem
\[
  \begin{aligned}
    \maximize_{\bm{P}} &  \text{Tr} \left( \bm{P}^H \bm{C}_{yx} \bm{C}_{xx}^{-1} \bm{C}_{xy} \bm{P} \right) \\
    \subjectto  & \text{rank } \bm{P} = r, \\
    & \bm{P}^H \bm{P} = \bm{I}.
  \end{aligned}
\]
The \( \bm{P} \) matrix is thus formed by the first \( r \) eigenvectors of the symmetric positive-definite matrix \( \bm{C}_{yx} \bm{C}_{xx}^{-1} \bm{C}_{xy} \).
Note that, if and only if \( \bm{X} \) is a full-rank \( m \times m \) matrix, these reduce to the first \( r \) left singular vectors of the output matrix \( \bm{Y} \) (i.e.\ the POD modes of \( \bm{Y} \)).
Once the \( \bm{P} \) matrix has been identified, the low-rank \( \bm{Q} \) matrix is given by
\[
  \bm{Q} = \bm{C}_{xx}^{-1} \bm{C}_{xy} \bm{P}
\]
Given \( \bm{P} \) and \( \bm{Q} \), the low-rank DMD operator can easily be factorized in a second step as
\[
  \bm{A} = \boldsymbol{\Uppsi \Uplambda \Upphi}^H,
\]
where \( \boldsymbol{\Uppsi}, \boldsymbol{\Upphi} \in \mathbb{C}^{m \times r} \) are the so-called left and right DMD eigenmodes, respectively, while \( \boldsymbol{\Uplambda} \in \mathbb{C}^{r \times r} \) is a diagonal matrix containing the DMD eigenvalues \( \lambda_i \).
These eigenvalues may provide valuable insights into the dynamics of the spatio-temporal coherent structures captured by the DMD eigenmodes.
A complete derivation of the optimal solution and of the equivalence between the norm-minimization and trace-maximization problems is provided in Appendix~\ref{appendix: dmd}.
It must be noted that most of the DMD variants proposed in the literature~\cite{jfm:rowley:2009,jfm:schmid:2010,jcd:tu:2014,pof:jovanovic:2014,jns:williams:2015,book:kutz:2016,expfluids:dawson:2016,jfm:noack:2016,siam:leclainche:2017,arxiv:hirsh:2019} actually compute only suboptimal approximations of the solution to the rank-constrained DMD problem based on various heuristics.
To the best of our knowledge, only H\'eas \& Herzet~\cite{arxiv:heas:2016} have proposed an alternative derivation of the optimal DMD solution albeit based on more advanced mathematics than the ones used herein.
Note finally that, using an information theoretic point of view, it has been shown by Tegmark~\cite{arxiv:tegmark:2019} for linear time-invariant stochastic dynamical systems that using two low-dimensional latent representations (namely \( \bm{P} \) and \( \bm{Q} \) or \( \boldsymbol{\Uppsi} \) and \( \boldsymbol{\Upphi} \)) enables us to better approximate the causal relationship between \( \bm{q}_k \) and \( \bm{q}_{k+1} \) than when using a single low-dimensional one as done in classical versions of DMD or when using POD to characterize dynamics rather than statistics.

\subsection{Sparse identification of nonlinear dynamics}\label{subsec: numerics -- sindy}

Identifying dynamical systems from data has been a central challenge in mathematical physics.
The form of the dynamics is typically either constrained via prior knowledge, as in Galerkin projection, or a particular model structure is chosen heuristically and parameters are optimized to match the data.
Simultaneous identification of the model structure and parameters is considerably more challenging as there are combinatorially many possible model structures.
The \emph{sparse identification of nonlinear dynamics} (SINDy) approach~\cite{pnas:brunton:2016} bypasses the intractable brute force search by leveraging the fact that many systems may be modeled as
\[
  \displaystyle \frac{\mathrm{d}\bm{x}}{\mathrm{d}t} =
  \bm{f}\left(\bm{x}\right)
\]
where \( \bm{x} \in \mathbb{R}^n \) is the state vector of our system and the unknown \( \bm{f} : \mathbb{R}^n \to \mathbb{R}^n \) is sparse in the space of possible right-hand side functions.

In order to identify \( \bm{f}\left(\bm{x}\right) \), time-series data is first collected and formed into the data matrix
\[
  \bm{X} = \begin{bmatrix} \bm{x}(t_1) & \bm{x}(t_2) & \cdots &
    \bm{x}(t_m) \end{bmatrix}^T.
\]
A similar matrix \(\dot{\bm{X}}\) of time derivatives is formed as well.
As a second step, a possibly over-complete library (or dictionary) \(\boldsymbol{\Uptheta}\) of candidate nonlinear functions is constructed, e.g.\
\[
  \boldsymbol{\Uptheta}\left(\bm{X}\right) = \begin{bmatrix} \bm{1} &
    \bm{X} & P_2(\bm{X}) & \cdots & P_d(\bm{X}) \end{bmatrix}.
\]
Here \(P_d(\bm{X})\) denotes a matrix with columns vectors given by all possible time-series of d\textsuperscript{th} degree monomials in the state \(\bm{x}\), e.g.\
\[
  P_2(x_1, x_2, x_3) = \begin{bmatrix} x_1^2 & x_1 x_2 & x_1 x_3 &
    x_2^2 & x_2 x_3 & x_3^2 \end{bmatrix}.
\]
Note that any basis function may be included in the library \(\boldsymbol{\Uptheta}\), albeit polynomials have proven to work well for fluid problems.
The unknown dynamical system may now be represented in terms of the data matrices as
\[
  \dot{\bm{X}} = \boldsymbol{\Uptheta}(\bm{X}) \boldsymbol{\Upxi}.
\]
Each column \(\boldsymbol{\Upxi}_k\) is a vector of coefficients determining the active terms in the equation governing the dynamics of \(x_k(t)\).
A parsimonious model will provide an accurate model fit with as few non-zero terms as possible in \(\boldsymbol{\Upxi}\).
Identifying this sparsity pattern may be formulated as the following optimization problem
\[
  \begin{aligned}
    \minimize_{\boldsymbol{\Upxi}_k} & \textrm{card}(\boldsymbol{\Upxi}_k) \\
    \subjectto & \| \dot{\bm{X}} - \boldsymbol{\Uptheta}(\bm{X}) \boldsymbol{\Upxi}_k \|_2^2 \leq \sigma \\
    & \bm{C} \boldsymbol{\Upxi}_k = \bm{d} \\
    & \bm{h}(\boldsymbol{\Upxi}_k) \leq 0,
  \end{aligned}
\]
where \(\textrm{card}(\boldsymbol{\Upxi}_k) = \| \boldsymbol{\Upxi}_k \|_0 \) is the cardinality (or \(\ell_0\) norm) of \(\boldsymbol{\Upxi}_k\), i.e.\ its number of non-zero entries, and the constraint \(\| \dot{\bm{X}} - \boldsymbol{\Uptheta}(\bm{X}) \boldsymbol{\Upxi}_k \|_2^2 \leq \sigma\) quantifies the fidelity of the model with respect to the data.
The linear equality constraint \( \bm{C} \boldsymbol{\Upxi}_k = \bm{d}\) and the convex inequality constraint \(\bm{h}(\boldsymbol{\Upxi}_k) \leq 0\) may be used to enforce additional constraints on the identified model based on prior physical knowledge such as energy-preserving quadratic non linearity in~\cite{jfm:loiseau:2018a} for instance.
The above minimization problem remains however a combinatorially complex optimization problem.
Various convex relaxations to this problem have been proposed over the years, most of them replacing the \( \ell_0 \) norm with an \( \ell_1 \) norm such as in LASSO regression~\cite{Tibshirani1996lasso}.
Due to the extended set of constraints that will be used in this work, the convex relaxation of this problem based on Thikonov regularization and iteratively thresholded least-squares was found to be sufficient.


\section{Results}\label{sec: results}

This section provides a factual presentation of the results.
First, results from the dimensionality reduction using dynamic mode decomposition are presented in \textsection~\ref{subsec: results -- low-dimensional embedding} while \textsection~\ref{subsec: results -- system identification} focuses on the identification of the low-order model governing the chaotic dynamics of the thermosyphon.

\subsection{Low-dimensional embedding}\label{subsec: results -- low-dimensional embedding}

The Navier-Stokes equations are integrated forward in time until a statistical steady state is reached.
Once reached, a sequence of 10 000 snapshots of the complete state vector of the fluctuation is collected with a sampling period \(\Delta T = 0.003\) diffusive time units.
Such a sampling period enables us to properly sample each oscillation of the flow in the clockwise or counter-clockwise direction using approximately 40 snapshots.
It must be noted moreover that, as shown in figure~\ref{fig: flow configuration -- example flow}, the flow exhibits symmetry with respect to the vertical axis.
By taking the symmetric of each collected snapshots, this dataset is thus artificially augmented to 20 000 snapshots.
This process enables us to enforce the existing spatial symmetry of the high-dimensional system directly in the training dataset itself.

\begin{figure}
  \centering
  \includegraphics[width=\textwidth]{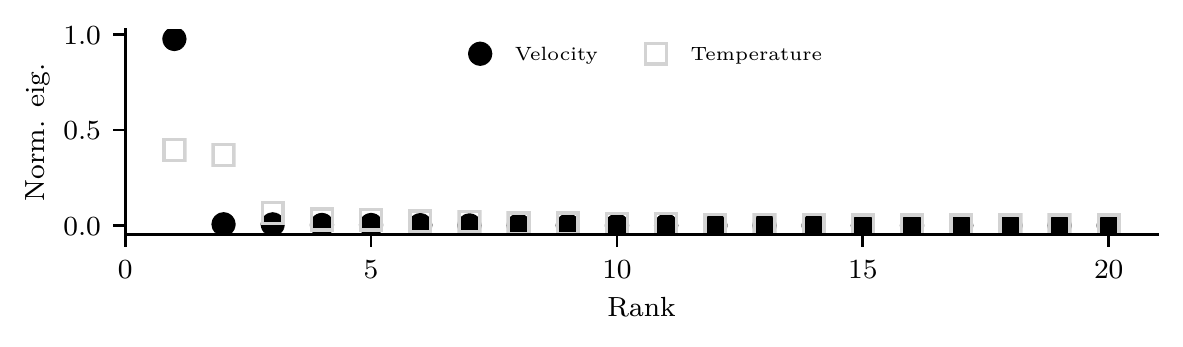}
  \caption{
    Eigenspectrum of the matrix \( \bm{C}_{yx} \bm{C}_{xx}^{-1} \bm{C}_{xy} \) for the velocity (blue circles) and the temperature (gray squares).
    For the velocity field, a rank 1 DMD model captures 98\% of the fluctuation's kinetic energy.
    For the temperature field, a rank 2 DMD mode captures 76\% of the fluctuation's thermal energy.
    For the latter, a DMD model of rank 20 would be needed to capture 99\% of the thermal energy.
  }\label{fig: results -- dmd selection}
\end{figure}

\begin{figure}
  \centering \includegraphics[width=\textwidth]{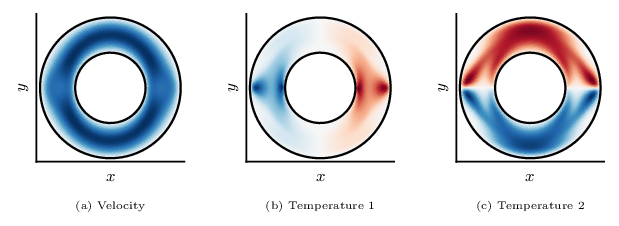}
  \caption{
    Spatial distribution of the DMD modes retained for our reduced-order model.
    Projection of the dynamics onto the span of these DMD modes is shown in figure~\ref{fig: results -- low-dimensional embedding}.
  }\label{fig: results -- dmd modes}
\end{figure}

Dedicated DMD analyses are conducted for the velocity and the temperature field separately.
Figure~\ref{fig: results -- dmd selection} depicts the eigenspectrum of the matrix \( \bm{C}_{yx} \bm{C}_{xx}^{-1} \bm{C}_{xy} \) for both fields.
It can be seen that, as for the Lorenz system, the velocity field is well approximated by a rank 1 model capturing close to 98\% of the fluctuation's kinetic energy.
Similarly, the temperature field is well approximated by a rank 2 model.
Note however that in this case the rank 2 model only captures 76\% of the fluctuation's thermal energy.
Capturing 98\% of the fluctuation's thermal energy would require a model of rank 20, thus significantly increasing the complexity of our overall modeling procedure.
The spatial distribution of these three DMD modes are shown in figure~\ref{fig: results -- dmd modes}.
While the DMD mode associated to the velocity field is almost invariant in the azimuthal direction, the modes associated to the temperature are anti-symmetric with respect to the vertical or horizontal axis.
Interestingly, the features captured by these DMD modes (namely the flow rate, the left-right and top-bottom temperature differences) correspond to the measurements considered by Erhard \& M\"uller~\cite{jfm:ehrhard:1990} to characterize experimentally the chaotic dynamics in a similar thermosyphon.
They are moreover consistent with the driving modes derived analytically by Yorke \emph{et al.}~\cite{physd:yorke:1987} using a Bessel-Fourier decomposition for a simplified set of partial differential equations modeling the same problem.

\begin{figure}
  \centering
  \includegraphics[width=\textwidth]{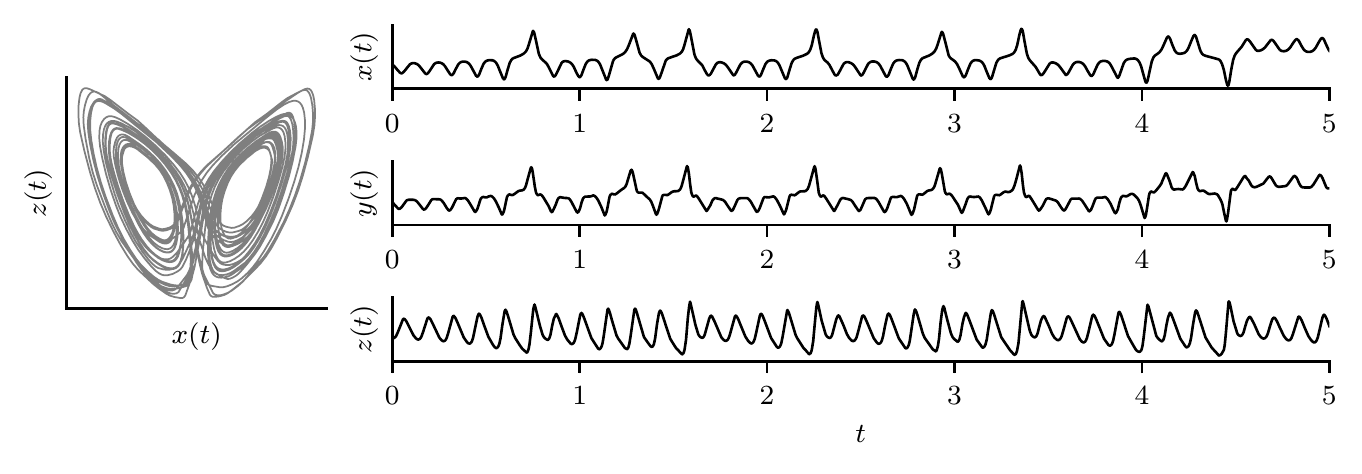}
  \caption{
    Low-dimensional embedding obtained by projecting time-series of the high-dimensional state vector onto the span of the chosen DMD modes.
    \( x(t) \) denotes the amplitude of the DMD mode associated to velocity while \( y(t) \) and \( z(t) \) are the amplitudes of the first and second DMD modes associated to the temperature field, respectively.
    Note that, for the sake of simplicity later on, time-series of \(x(t)\), \(y(t)\) and \(z(t)\) have been standardized such that they all have zero-mean and unit-variance.
  }\label{fig: results -- low-dimensional embedding}
\end{figure}

Figure~\ref{fig: results -- low-dimensional embedding} shows the projection of time-series of the high-dimensional state vector onto the span of these DMD modes.
As we had expected, the attractor within this low-dimensional space is strikingly similar to the canonical Lorenz attractor, see figure~\ref{fig: appendix -- lorenz attractor} for a visual comparison.
Although not shown, it has the same \((x, y, z) \mapsto (-x, -y, z) \) equivariance as the canonical Lorenz attractor.
These time-series will form the training dataset for the identification of the low-order model in the next section.

\subsection{System identification}\label{subsec: results -- system identification}

Let us now turn our attention to the identification of a low-order model able to approximate correctly the dynamics of the DMD measurements presented in the previous section.
Note that the amplitudes \( \bm{x}(t) \) of the DMD modes are mean-centered and normalized to unit-variance to ease the identification process.
Additionally, we will assume that the unknown system exhibits quadratic nonlinearities (consistent with the Navier-Stokes equations) so that the library \( \boldsymbol{\Uptheta}(\bm{x}) \) of candidate functions for the identification is given by
\[
  \boldsymbol{\Uptheta}(x, y, z) =
  \begin{bmatrix}
    1 & x & y & z & x^2 & xy & xz & y^2 & yz & z^2
  \end{bmatrix}.
\]
Under these assumptions, our unknown model takes the general form
\[
  \begin{aligned}
    \dot{x} & = a_0 + a_1 x + a_2 y + a_3 z + a_4 x^2 + a_5 xy + a_6 xz + a_7 y^2 + a_8 yz + a_9 z^2, \\
    \dot{y} & = b_0 + b_1 x + b_2 y + b_3 z + b_4 x^2 + b_5 xy + b_6 xz + b_7 y^2 + b_8 yz + b_9 z^2, \\
    \dot{z} & = c_0 + c_1 x + c_2 y + c_3 z + c_4 x^2 + c_5 xy +
    c_6 xz + c_7 y^2 + c_8 yz + c_9 z^2.
  \end{aligned}
\]
With no other assumptions about the dynamics, up to thirty coefficients thus need to be identified from our training dataset.

\subsubsection{Deriving the constraints}\label{subsubsec: results -- deriving the constraints}

Assuming that the low-order dynamics are governed by a Lorenz-like process, \textsection~\ref{subsec: lorenz -- properties} has introduced the various properties the model would have to exhibit.
Because our unknown model is linear-in-parameters, these properties can easily be enforced in the identification step by constraining the coefficients of the model.

\paragraph{Equivariant dynamics :} The first property we will discuss is that of equivariance.
Analysis of the phase portraits shows that the dynamics of the thermosyphon embedded into the low-dimensional DMD space appear to be equivariant with respect to the change of coordinates
\[
  (x, y, z) \mapsto (-x, -y, z).
\]
Denoting by
\[
  \boldsymbol{\gamma} =
  \begin{bmatrix}
    -1 & 0 & 0 \\
    0 & -1 & 0 \\
    0 & 0 & 1
  \end{bmatrix}
\]
the operator performing this change of coordinates, the yet-to-be identified unknown system thus needs to satisfy
\[
  \boldsymbol{\gamma} \dot{\bm{x}} = \bm{f}\left(
    \boldsymbol{\gamma} \bm{x} \right).
\]
for the dynamics to be equivariant.
Given our choice for the library \( \boldsymbol{\Uptheta}(x, y, z) \), a dynamical system equivariant under the action of \( \boldsymbol{\gamma} \) would take the form
\[
  \begin{aligned} 
    \dot{x} & = a_0 x + a_1 y + a_3 xz + a_4 yz, \\
    \dot{y} & = b_0 x + b_1 y + b_3 xz + b_4 yz, \\
    \dot{z} & = c_0 + c_1 z + c_2 x^2 + c_3 xy + c_4 y^2 + c_5
    z^2.
  \end{aligned}
\]
Explicitly enforcing the equivariance constraint into the structure of the model thus reduces the number of coefficients to be identified from thirty down to only fourteen.

\paragraph{Dissipative dynamics :} The second property that needs to be verified is that of dissipative dynamics.
As shown in \textsection~\ref{subsec: lorenz -- properties}, this property reads
\[
  \nabla \cdot \bm{f}\left( \bm{x} \right) < 0 \quad \forall
  \bm{x}.
\]
Starting from the equivariant model, the divergence of the unknown right-hand side function is given by
\[
  \nabla \cdot \bm{f}\left( \bm{x} \right) = a_0 + b_1 + c_1 +
  \left(a_3 + b_4 + c_5 \right) z.
\]
In order for our model to have dissipative dynamics in the whole phase space, the unknown coefficients must satisfy the following constraints
\[
  \begin{aligned}
    a_0 + b_1 + c_1 & < 0 \\
    a_3 + b_4 + 2 c_5 & = 0.
  \end{aligned}
\]
Note that this set of constraints couples all of the dynamical equations thus forcing us to identify all of them at once rather than sequentially.

\paragraph{Energy-preserving nonlinearities : } The quadratic nonlinearities of the incompressible Navier-Stokes equations do not create nor dissipate energy but simply scatters it among the different length scales.
As shown in~\cite{jfm:loiseau:2018a} this property can also be enforced in the identification process using linear equality constraints.
Starting from the equivariant model, the equation governing the energy \( E = \nicefrac{x^2 + y^2 + z^2}{2} \) needs to read
\[
  \dot{E} = c_0 z + a_0 x^2 + b_1 y^2 + c_1 z^2 + (a_1 + b_0) x y
\]
in order to ensure that the quadratic nonlinearity of the unknown model is energy-preserving.
As a consequence, the coefficients of the equivariant model need to satisfy
\[
  \begin{aligned}
    b_3 + c_4 & = 0, \\
    a_2 + c_2 & = 0, \\
    a_3 + b_2 + c_3 & = 0 \\
    c_5 & = 0.
  \end{aligned}
\]

\paragraph{Stable manifold of the conducting state :} The last property discussed in \textsection~\ref{subsec: lorenz -- properties} stated that the \(z\)-axis belongs to the stable manifold of the fixed point associated to the pure conducting state.
Assuming that \( x = y = 0 \), the equation for \(z\) simply reads
\[
  \dot{z} = c_0 + c_1 z.
\]
The corresponding fixed point is given by \( z = -\nicefrac{c_0}{c_1} \).
Additionally, this fixed point is stable provided that
\[
  c_1 < 0.
\]
This linear inequality constraint is the last constraint we will enforce in the identification process.

\subsubsection{Identifying the system}

\paragraph{Convex relaxation :} As discussed in \textsection~\ref{subsec: numerics -- sindy}, the optimization problem to be solved for the identification is technically a combinatorial problem which becomes rapidly intractable as the number of candidate functions in the library \( \boldsymbol{\Uptheta}(\bm{x}) \) increases.
In order to bypass this issue, a convex relaxation of the problem based on Thikonov regularization is used in the present work.
The corresponding constrained convex minimization problem then reads
\begin{equation}
  \begin{aligned}
    \minimize_{\boldsymbol{\Upxi}} & \sum_{k=1}^3 \| \boldsymbol{\Uptheta}(\bm{x}) \boldsymbol{\upxi}_k - \dot{\bm{x}}_k \|_2^2 + \lambda \| \boldsymbol{\upxi}_k \|_2^2 \\
    \subjectto & \bm{C} \boldsymbol{\Upxi} = 0 \\
    & \bm{H} \boldsymbol{\Upxi} \leq 0.
  \end{aligned}
  \label{eq: results -- convex min problem}
\end{equation}
with \( \boldsymbol{\Upxi} = \begin{bmatrix} \boldsymbol{\upxi}_1 & \boldsymbol{\upxi}_2 & \boldsymbol{\upxi}_3 \end{bmatrix}^T \) the vector of unknown coefficients for the three equations and \( \lambda \) a hyper parameter to be optimized.
Note that the \( \ell_2 \)-regularization used herein differs from the sparsity-promoting approach classically used in SINDy~\cite{pnas:brunton:2016,ieee:mangan:2016,ifac:brunton:2016,nature:brunton:2017,sciadv:rudy:2017,arxiv:kaiser:2017,jfm:loiseau:2018a,jfm:loiseau:2018b}.
Although Thikonov regularization is not per say a sparsity promoting regularization, the admissible structure of the model is already sufficiently restricted by the physics-derived constraints discussed in the previous section and this simple approach was found to be sufficient.

\begin{figure}
  \centering
  \includegraphics[width=\textwidth]{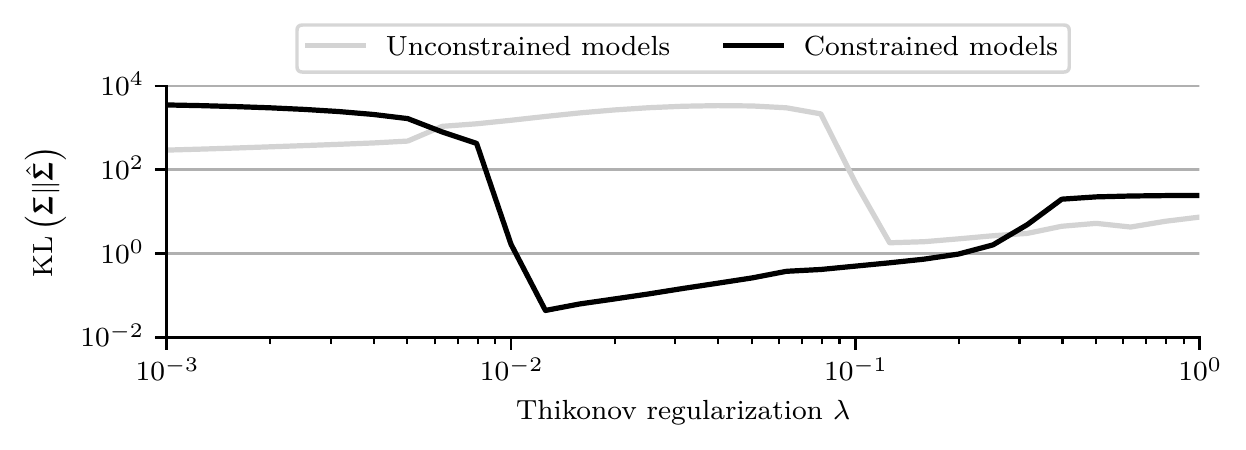}
  \caption{
    Evolution of the Kullback-Leibler divergence of the variance-covariance matrices as a function of the regularization parameter \( \lambda \).
    Unconstrained models correspond to the most general models with up to 30 coefficients that need to be identified without prior physical knowledge.
    Constrained models correspond to models identified using all of the physics constraints discussed in \textsection~\ref{subsubsec: results -- deriving the constraints}.
  }\label{fig: results -- model selection}
\end{figure}

\paragraph{Model selection :} Solving the optimization problem~\eqref{eq: results -- convex min problem} gives rise to a family of candidate models parameterized by \( \lambda \) from which a single one needs to be selected.
To do so, each low-order model from this family is integrated forward in time until a statistical steady state is reached and the resulting empirical variance-covariance matrix \( \hat{\boldsymbol{\Upsigma}} \) is compared against the variance-covariance matrix \( \boldsymbol{\Upsigma} \) obtained from the training dataset.
This comparison relies on the Kullback-Leibler divergence between two symmetric positive-definite matrices.
It is defined as
\[
  KL \left( \boldsymbol{\Upsigma} \| \hat{\boldsymbol{\Upsigma}} \right) = \frac{1}{2} \left( \mathrm{Tr}\left( \hat{\boldsymbol{\Upsigma}}^{-1} \boldsymbol{\Upsigma} \right) - n + \ln \frac{\vert \hat{\boldsymbol{\Upsigma}} \vert}{\vert \boldsymbol{\Upsigma} \vert} \right)
\]
where \( n \) is the number of degrees of freedom (here 3) and \( \vert \boldsymbol{\Upsigma} \vert \) denotes the determinant of said matrix.
Figure~\ref{fig: results -- model selection} shows the evolution of this metric as a function of the regularization parameter \( \lambda \) for the two classes of models, namely unconstrained and physics-constrained models.
Enforcing prior physical knowledge into the identification procedure leads to orders of magnitude more accurate models.

\begin{figure}
  \centering
  \includegraphics[width=\textwidth]{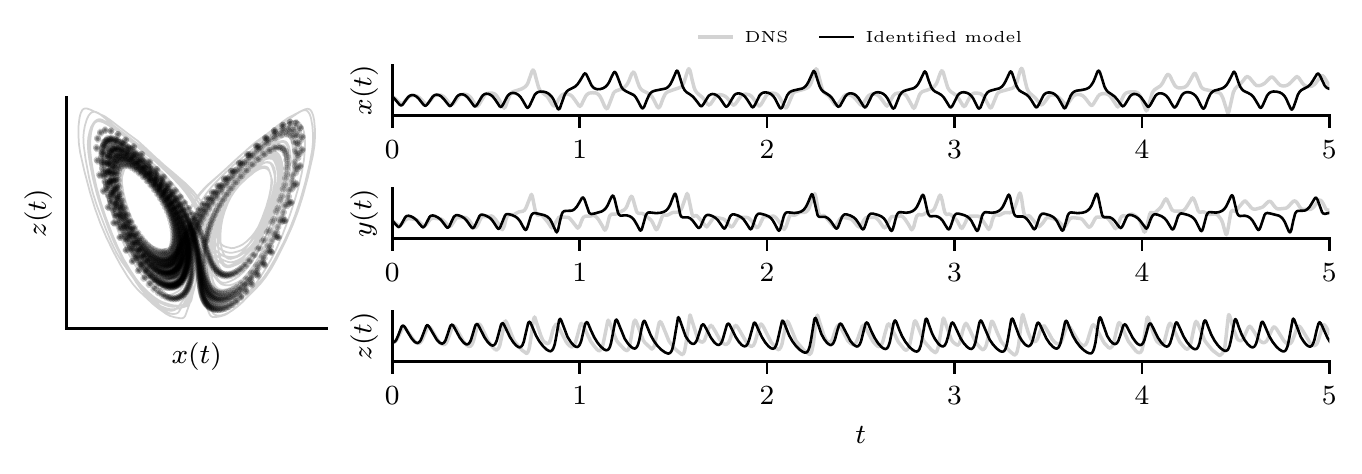}
  \caption{
    Comparison of the dynamics of the chaotic thermosyphon embedded in the low-dimensional DMD space (light gray) and those of the identified low-order model (black).
    Left panel shows a projection of the Lorenz-like strange attractor onto the \((x, z)\) plane.
    Right panel shows typical time series of the different degrees of freedom.
  }
  \label{fig: results -- identified model}
\end{figure}

\paragraph{Identified model :} Based on figure~\ref{fig: results -- model selection}, the constrained model obtained for \( \lambda \simeq 10^{-2} \) is considered the best model.
Its coefficients are then refitted using an extended least squares procedure~\cite{book:billings:2013} to remove any left over correlation in the residuals in order to obtain unbiased estimates.
Rounding its coefficients to two significant digits, this model reads
\[
  \begin{aligned}
    \dot{x} & = -76.08x + 88.39y \\
    \dot{y} & = 20.76x - 41.49 xz - 4.19 y \\
    \dot{z} & = 41.49 xy - 43.67 - 17.31z.
  \end{aligned}
\]
By design, it verifies all of the physical constraints discussed in \textsection~\ref{subsubsec: results -- deriving the constraints}.
Except for the constant term in the \( z \)-equation, the structure of this model is moreover identical to that of the Lorenz system, see \textsection~\ref{sec: lorenz}.
Figure~\ref{fig: results -- identified model} provides a qualitative comparison of the true strange attractor and the one of the identified model along with representative time series of the various degrees of freedom.
Good agreement in the ``eye-ball norm'' can be observed.
More quantitative analyses and discussion about the accuracy of the identified model are postponed to \textsection~\ref{subsec: discussion -- low-order model}.

\section{Discussion}\label{sec: discussion}

After having presented the results in \textsection~\ref{sec: results}, this section provides a more in-depth discussion about the model's properties as well as its interpretation in terms of physical mechanisms.
Discussion pertaining to dynamic mode decomposition is presented in \textsection~\ref{subsec: discussion -- dmd} while properties of the identified low-order model are discussed in \textsection~\ref{subsec: discussion -- low-order model}.

\subsection{On the dynamic mode decomposition}\label{subsec: discussion -- dmd}

\begin{figure}
  \centering
  \subfigure[]{\includegraphics[width=\textwidth]{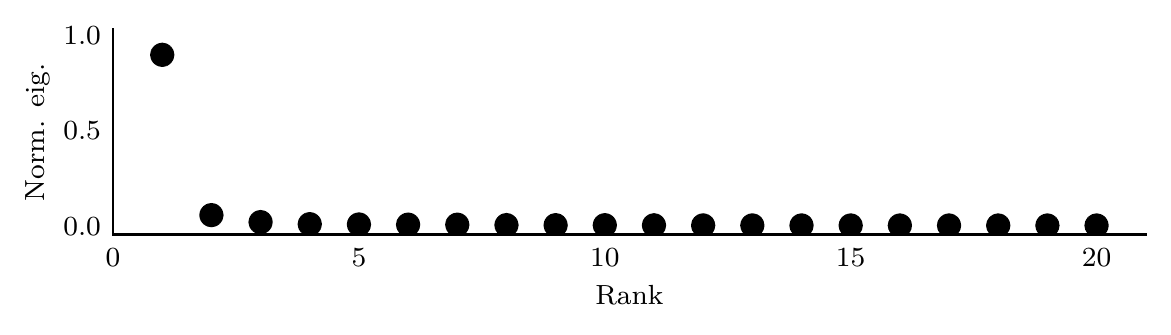}} \\
  \subfigure[]{\includegraphics[width=\textwidth]{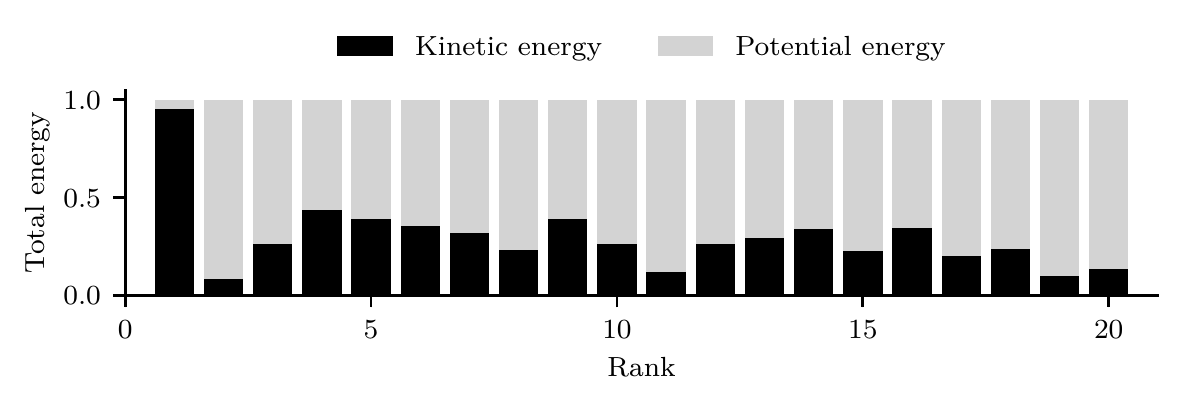}}
  \caption{
    (a) Eigenspectrum of the matrix \( \bm{C}_{yx} \bm{C}_{xx}^{-1} \bm{C}_{xy} \) from the DMD optimization problem when the embedding for the velocity and the temperature is learned jointly using the inner product~\eqref{eq: discussion -- dmd full norm}.
    (b) Energy budget (i.e. fraction of the kinetic and thermal energy forming the total energy) for the leading eigenvectors of the matrix \( \bm{C}_{yx} \bm{C}_{xx}^{-1} \bm{C}_{xy} \).
    Note that the DMD modes considered in this section are a linear combination of the first three eigenvectors.
  }
  \label{fig: discussion -- dmd full selection}
\end{figure}

In this work, choice has been made to learn a low-dimensional embedding for the velocity and temperature fields separately.
Based on the eigenvalue distribution of symmetric positive definite matrices, a rank 1 model was found to be sufficient for the velocity field while a rank 2 model (albeit less accurate) was deemed enough for the temperature field.
Following \cite{pof:john:2012}, one could have alternatively defined the inner product inducing the total energy as
\begin{equation}
  \langle \bm{q}_i, \bm{q}_j \rangle = \frac{1}{2} \int_{\Omega} \bm{u}_i \cdot \bm{u}_j + \gamma \boldsymbol{\theta}_i \cdot \boldsymbol{\theta}_j \ \mathrm{d}\Omega
  \label{eq: discussion -- dmd full norm}
\end{equation}
and learn the embedding for the complete state vector at once.
Here, \( \gamma \) weights the kinetic energy and the thermal one.
Depending on its value, different embeddings could be obtained.
When \( \gamma \) is set to \( Ra \ Pr \) and \( i = j \), the first term describes the kinetic energy of the fluctuation while the second one corresponds to its non-dimensional potential energy.
Figure~\ref{fig: discussion -- dmd full selection} (a) shows the eigenvalue distribution of the symmetric positive definite matrix \( \bm{C}_{\bm{yx}} \bm{C}_{\bm{xx}}^{-1} \bm{C}_{\bm{xy}}\) when the whole state vector is considered at once for \( \gamma = Ra \ Pr \) while figure~\ref{fig: discussion -- dmd full selection} (b) shows how the total energy is distributed between kinetic and potential energy for the leading eigenvectors of this symmetric positive definite matrix.
Looking at these two figures, one can observe that a rank 1 model captures close to 95\% of the linearly predictable evolution of the fluctuation's total energy.
Additionally, almost all of this total energy consists in kinetic energy.
On the other hand, the total energy of the second leading eigenvector mainly consists in potential energy.
This clear distinction between the distribution of energy for the first two eigenvectors supports our decision to learn a separate embedding for the velocity and the temperature.

\begin{figure}
  \centering
  \subfigure[DMD modes obtained using the total energy inducing inner product]{\includegraphics[width=\textwidth]{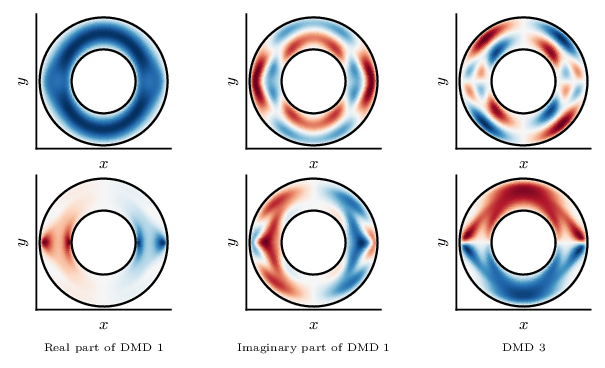}}
  \subfigure[Embedding for the velocity and temperature learned jointly]{\includegraphics[width=\textwidth]{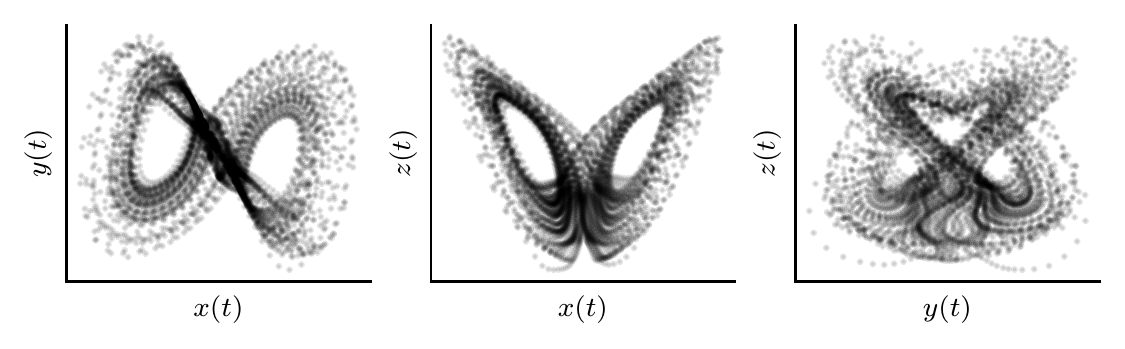}}  
  \subfigure[Embedding for the velocity and temperature learned separately]{\includegraphics[width=\textwidth]{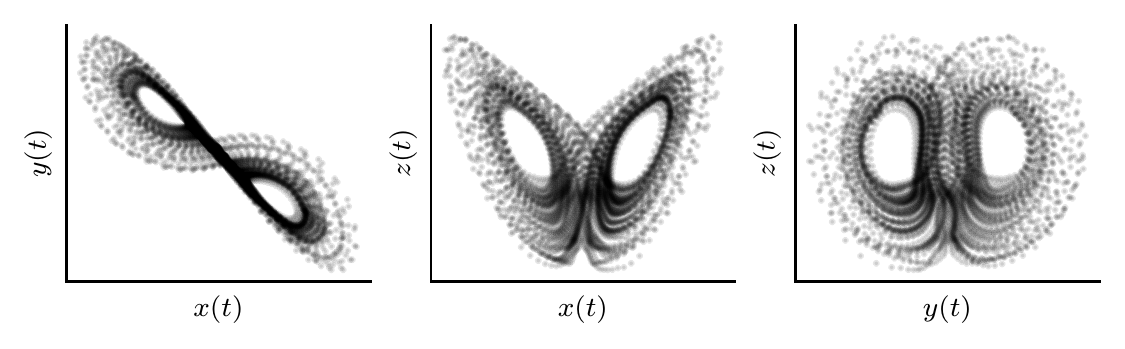}}
  \caption{
    (a) Azimuthal velocity (top row) and temperature (bottom row) fields of the DMD modes associated to the rank 3 model obtained using the inner product inducing the total energy.
    The first two columns correspond to the real and imaginary part of the complex-conjugate DMD mode while the last column corresponds to the real-valued DMD mode.
    To be compared with figure~\ref{fig: results -- dmd modes}.
    (b) Low-dimensional representation obtained after projecting the training dataset onto the span of the modes above.
    (c) Low-dimensional representation used in this work obtained after learning the embedding for the velocity and the temperature separately.
  }\label{fig: discussion -- full dmd embedding}
\end{figure}

Figure~\ref{fig: discussion -- full dmd embedding} (a) depicts the azimuthal velocity component and temperature field of the DMD modes when a rank 3 model is considered while figures~\ref{fig: discussion -- full dmd embedding} (b) and (c) provides a comparison of the low-dimensional representation obtained from learning jointly or separately the embedding of the state vector, respectively.
Because they are highly linearly correlated, the real part of the first DMD mode includes both the leading coherent structure associated to the velocity and the left-right temperature fluctuation.
As a consequence, these two structures are fused into a single degree of freedom when the embedding for the velocity and the temperature is learned jointly.
With the exception of its velocity component, the third DMD mode captures as before the leading coherent structure associated to the vertical temperature fluctuation.
Figures~\ref{fig: discussion -- full dmd embedding}(b) and (c) provide a comparison of the resulting low-dimensional representations.
While the projection onto the \((x, z)\) plane appears similar for both embeddings, the projection onto the other planes are different.
Despite our efforts, we have not been able to successfully identify a low-order model when learning the embedding for the velocity and temperature jointly.
This failed attempt highlights the importance of choosing judiciously the embedding when aiming for the identification of an interpretable low-order model and seems to suggest that learning a low-dimensional embedding for each physical quantity separately might provide more useful information into the dynamics of the system than when learning the embedding for the whole state vector at once.
This might be better understood by realizing that, given the state vector \( \bm{q} = \begin{bmatrix} \bm{v} & \boldsymbol{\theta} \end{bmatrix}^T \), the DMD operator may be partitionned as
\[
  \begin{bmatrix} \bm{v}_{k+1} \\ \boldsymbol{\theta}_{k+1} \end{bmatrix}
  =
  \begin{bmatrix}
    \bm{A}_{\bm{vv}} & \bm{A}_{\bm{v} \boldsymbol{\theta}} \\
    \bm{A}_{\boldsymbol{\theta}\bm{v}} & \bm{A}_{\boldsymbol{\theta \theta}}
  \end{bmatrix}
  \begin{bmatrix} \bm{v}_{k} \\ \boldsymbol{\theta}_{k} \end{bmatrix}.
\]
Performing a component-based DMD analysis thus enables us to obtain different (and possibly more accurate) low-rank approximations of each block while also providing a finer-grain description of the interplay between velocity and temperature.
To date, this observation however lack theoretical justifications.

\subsection{On the identified low-order model}\label{subsec: discussion -- low-order model}

\subsubsection{How accurate is it?}\label{subsubsec: discussion -- low-order model accuracy}

\begin{figure}
  \centering
  \includegraphics[width=\textwidth]{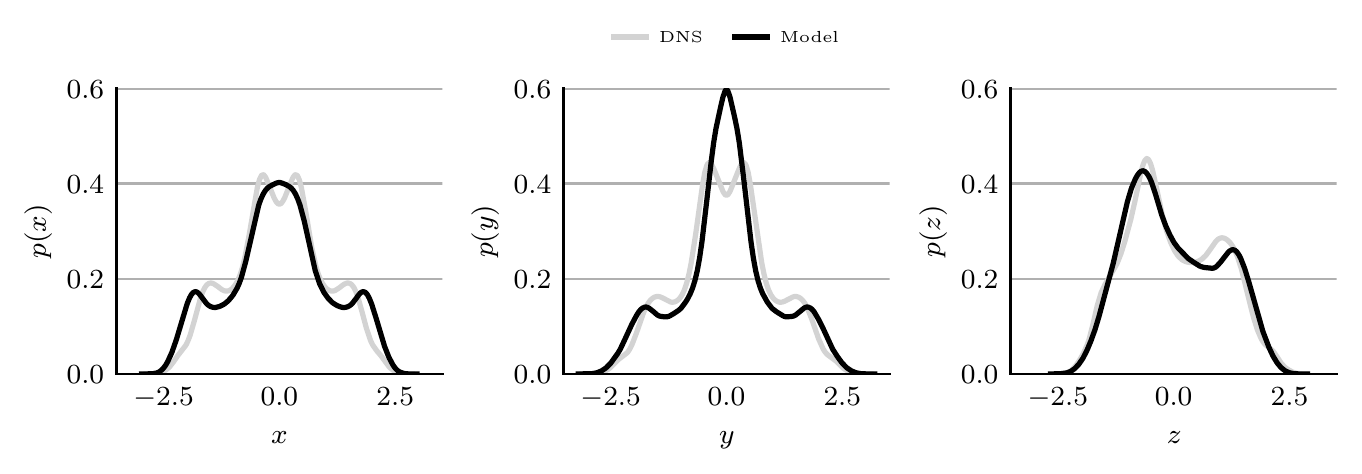}
  \caption{
    Comparison of the probability density function of the different degrees of freedom.
    For the ground truth (light gray), these pdf have been computed from the projection of a test dataset onto the span of the DMD modes.
    For the ones predicted by the model (black), these have been computed after simulating the identified low-order model over 650 non-dimensional time units.
  }\label{fig: discussion -- pdf comparisons}
\end{figure}

\begin{figure}
  \centering
  \subfigure[Autocorrelation function]{\includegraphics[width=.495\textwidth]{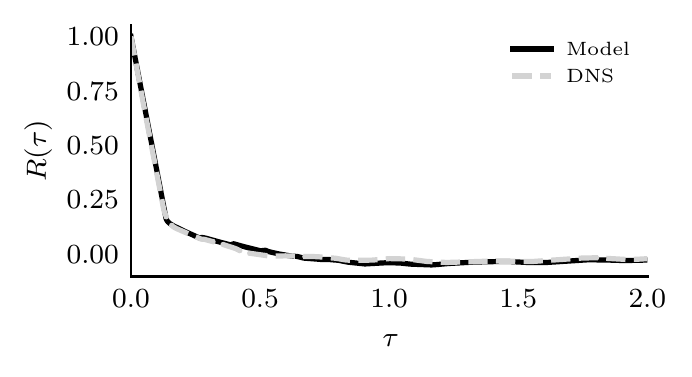}}%
  \hfill
  \subfigure[Time-scale distribution]{\includegraphics[width=.495\textwidth]{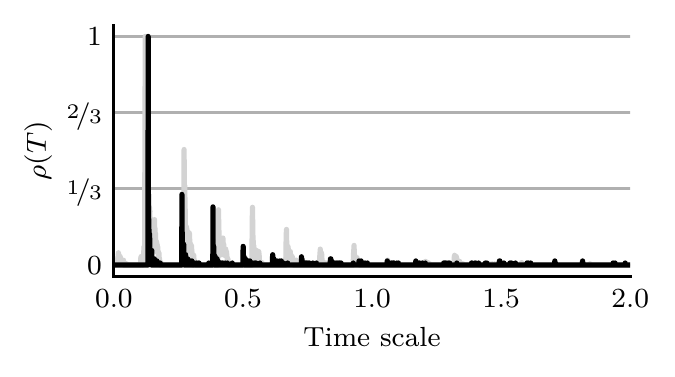}}
  \caption{
    Comparison of the autocorrelation function and time-scale distributions obtained from direct numerical simulation (light gray) and the low-order model (black).
    For the direct numerical simulation, these statistical quantities have been computed using the time-series of the flow rate while that of \( x(t) \) (associated to the velocity DMD mode) has been used for the low-order model.
    Both autocorrelation functions have been computed for the corresponding telegraph signal.
  }\label{fig: discussion -- statistics comparison}
\end{figure}

While figure~\ref{fig: results -- identified model} shows good qualitative agreements between the dynamics of the thermosyphon embedded within the low-dimensional DMD space and that of the identified model, let us now try to characterize this agreement more quantitatively.
Figure~\ref{fig: discussion -- pdf comparisons} provides a comparison of the probability distribution functions obtained from direct numerical simulation of the Navier-Stokes equations and those obtained from the identified low-order model.
Given that the low-order model only has three degrees of freedom, excellent agreement is obtained for all three probability density functions.
A small discrepancy is however observed for small values of \( x(t) \) and \( y(t) \).
It corresponds to situations where the flow is about to switch from clockwise to anti-clockwise direction (or vice-versa).
During these events, the amplitude of the DMD modes associated to the velocity and the left-right temperature difference are close to zero.
Hence, a plausible explanation for this mismatch is that, during these events, the dynamics of the flow are dominated by other coherent structures not accounted for by the present model.
The overall dynamics of the system are nonetheless correctly captured as assessed by figure~\ref{fig: discussion -- statistics comparison}.
Using the same symbolic dynamics approach as in \textsection~\ref{subsec: flow configuration -- statistical analysis}, it compares (a) the autocorrelation function of the telegraph signal and (b) the time-scale distribution obtained from direct numerical simulation of the Navier-Stokes equations and by our low-order model.
Remarkable agreement is obtained for the autocorrelation function.
Reasonable agreement is also obtained regarding the time-scale distribution.
For all intent and purposes the low-order model identified in the present work thus captures remarkably accurately the statistical and dynamical properties of the flow.

\subsubsection{Interpretability and physical processes captured by the model}\label{subsubsec: discusssion -- low-order model physics}

Over the past few years, interpretability has been an overarching goal in machine learning, in particular when applied to modeling of physical systems.
Because it relies on the joint use of dynamic mode decomposition and SINDy, every single mode of the low-dimensional embedding and term of the low-order model identified in the present work can be understood as modeling a given coherent structure or physical mechanism at play in the thermosyphon.
Additionally, a sequence of linear change of variables and re-scaling time enables us to recast the identified model into the Lorenz system
\[
  \begin{aligned}
    \dot{x} & = \sigma ( y - x ) \\
    \dot{y} & = \rho x - y - xz \\
    \dot{z} & = xy - \beta z
  \end{aligned}
\]
with \( (\sigma, \rho, \beta) = (18.14, 34.75, 4.13) \), thus effectively reducing the number of parameters from seven down to only three and facilitating their interpretability and analysis.

Analysis of the above model shows that it has three fixed points.
The trivial one, given by \( (x, y, z) = (0, 0, 0) \), corresponds to the pure conducting state (i.e. only temperature diffusion and no net flow through a cross section of the thermosyphon).
As expected, this particular fixed point is moreover linearly unstable with its unstable eigenvector corresponding to the linear seed of a convection cell.
The other two fixed points are symmetric of one another and correspond to the clockwise and anti-clockwise configurations.
Interestingly, given the set of identified parameters, these fixed points appear to be very slightly stable, thus suggesting that the dynamics observed in the thermosyphon do not correspond to full-blown chaos but only transient chaos.
It is unclear at the present time whether this is an actual feature of the flow or only a limitation of the identification procedure used in this work.
It must be noted however that simulations of the identified model show that this transient chaos can nonetheless sustains itself for a period of time larger than hundred times the complete duration of the direct numerical simulation used to generate our training data.
Moreover, increasing the training data with another set of 10 000 snapshots and re-running the whole identification procedure leaves our results largely unchanged.
In the absence of further evidence, one cannot conclude about the nature of the dynamics (chaos or only transient chaos) of the Navier-Stokes equations for the set of physical parameters considered herein.

Because it reduces to the canonical Lorenz system, the various terms in the identified can be interpreted as modeling similar physical mechanisms as those observed in the canonical Rayleigh-B\'enard convection.
For instance, in the governing equation for \( x(t) \), the term \( - \sigma x \) serves to model momentum diffusion due viscous dissipation while \( \sigma y \) captures the influence of buoyancy on the flow.
Similarly for the equation governing \( y(t) \), the term \( -y \) captures the diffusion of left-right temperature fluctuation while the terms \( \rho x \) and \( - xz \) serve to model the advection of the mean temperature profile and vertical temperature fluctuation, respectively.
Finally, the terms \( xy \) and \( -\beta z \) in the last equation model the advection of the left-right temperature fluctuation and the diffusion of vertical temperature fluctuation.
Regarding the parameters of the model, it is well known from the Lorenz system that \( \rho \) is the ratio of the Rayleigh number \( Ra \) over the critical Rayleigh number \( Ra_c \) corresponding to the onset of the convection cells.
Similarly, the parameters \( \sigma \) and \( \beta \) can be directly related to the Prandtl number and the aspect ratio of the convection cells.
Given that no prior knowledge about the system other than the physical constraints it needs to satisfy had been included in the identification process, the value \( \rho = 34.75 \) is in excellent agreement with the expected ratio of \( Ra \) over \( Ra_c \).
Indeed, preliminary computations of the complete bifurcation diagram of the high-dimensional system shows that the critical Rayleigh number for the onset of the steady convection cells is \( Ra_c = 495 \).
For the Rayleigh number considered in this work, this would thus correspond to \( \rho = 34.34 \), i.e.\ only 1\% different than the value identified.
It is thus quite remarkable that a model identified at \( Ra = 17\ 000 \) is able to accurately predict a fundamental property of the high-dimensional such as a bifurcation happening at a fairly different value of the control parameter.
Note however that, since a single operating condition has been considered herein, this excellent agreement might be only coincidental and further analyses are required to confirm it.


\section{Conclusion and perspectives}\label{sec: conclusion}

Identifying accurate yet interpretable low-order model from data has been an overarching goal in mathematical physics which has gained a renewed interest over the past decade with the advances made in dimensionality reduction and sparsity-promoting regression techniques.
In this work, we have illustrated how such techniques, namely dynamic mode decomposition~\cite{jfm:schmid:2010} and SINDy~\cite{pnas:brunton:2016}, could be used to identify a low-order model of the chaotic thermal convection taking place in an annular thermosyphon at moderate Rayleigh numbers.
Particular emphasis has been put on the necessity for the low-dimensional embedding and the identified system to comply with prior physical knowledge.
To do so, the construction of the low-order model proposed herein closely follows the derivation of the canonical Lorenz system from the Oberbeck-Boussinesq equations modeling the Rayleigh-B\'enard convection between two infinite flat plates.
Given the simplicity of the identified model compared to the numerical simulation of the original high-dimensional system, excellent qualitative and quantitative agreements have been obtained as assessed by the similar structure of the strange attractor onto which the dynamics evolve in both situations as well as various other statistical quantities.

From a physical point of view, this work further confirms a strand of evidence~\cite{physd:yorke:1987,jfm:ehrhard:1990,ijhmt:louisos:2015} that the chaotic dynamics resulting from unstable temperature stratification in annular thermosyphons may be correctly captured using a low-order model as simple as the canonical Lorenz system.
Numerous questions however remain unanswered, the most important ones being
\begin{itemize}
\item Are the conclusions drawn from this study applicable to other configurations?
  Is the Lorenz model a universal low-order model for other unstable temperature stratifications (e.g.\ Rayleigh-B\'enard flows in rectangular enclosures~\cite{pof:podvin:2012,jfm:podvin:2019}) at low or moderate Rayleigh numbers?

\item To what extent does the Lorenz system, combined with a DMD embedding, provides a good approximation of the dynamics as the Rayleigh number increases?
  How should one modify the model in order to account for turbulence once it sets in?
\end{itemize}
While it may be relatively easy to provide an answer to the first questions by running the same analyses as those conducted herein for other geometries, the answer to the second questions is highly dependent on how compressible (from a dimensionality point of view) the dynamics of the high-dimensional system are once turbulence sets in.

Finally, it should be emphasized that our ability to identify low-order models from data stemming from a high-dimensional system with various practical applications opens new possibilities.
In particular, both dynamic mode decomposition and SINDy can easily be extended to include external input~\cite{siam:proctor:2016,ifac:brunton:2016}, enabling us to identify accurate low-order models for control purposes.
These models could moreover be used in conjunction with optimal sensor placement~\cite{ieee:manohar:2018,ieee:clark:2019} to enable real-time monitoring and estimation of the system.
Using tool from \emph{model predictive control}, such low-order models could be used to strengthen or decrease the chaotic dynamics depending on the applications.
A first step toward this goal has been achieved by Kaiser et al.~\cite{arxiv:kaiser:2017}.


\begin{acknowledgements}
  This work is adapted from a contribution to the long program Machine Learning for Physics and the Physics of Learning, organized by the Institute for Pure and Applied Mathematics (IPAM) at UCLA (Los Angeles, USA) in 2019 and made possible thanks to the financial support of IPAM, which is supported by the National Science Foundation (NSF Grant No.\ DMS-1440415).
  The author would also like to gratefully thank Steven Brunton, Jared Callahan, Kathleen Champion, Onofrio Semeraro, Alessandro Bucci and many others for the fruitful dicussions on this topic.
\end{acknowledgements}

\appendix
\section{Derivation of the optimal solution to the DMD
  problem}\label{appendix: dmd}

\subsection{Equivalence between norm-minimization and
  trace-maximization formulation of the DMD problem}

This section aims at deriving the equivalence between the
norm-minimization and trace-maximization formulations of the DMD
problem and the corresponding solution.  Given two data sequences
\( \bm{X} \) and \( \bm{Y} \) related by
\[
  \bm{y}_k = \bm{f}(\bm{x}_k),
\]
the aim of DMD is to find the low-rank operator \( \bm{A} \) that best
approximates the possibly nonlinear function \( \bm{f} \) in the
least-squares sense.  As discussed in \textsection~\ref{subsec:
  numerics -- dmd}, this can be formulated as the following
rank-constrained minimization problem
\begin{equation}
  \begin{aligned}
    \minimize_{\bm{P}, \bm{Q}} & \| \bm{Y} - \bm{PQ}^H \bm{X} \|_F^2 \\
    \subjectto & \text{rank } \bm{P} = \text{rank } \bm{Q} = r \\
    & \bm{P}^H \bm{P} = \bm{I},
  \end{aligned}
  \label{eq: appendix -- dmd norm minimization}
\end{equation}
where we use the fact the unknown low-rank operator can be factorized
as \( \bm{A} = \bm{PQ}^H \).  Introducing the cost function
\[
  \begin{aligned}
    \mathcal{J}(\bm{P}, \bm{Q}) & = \| \bm{Y} - \bm{PQ}^H \bm{X} \|_F^2 \\
    & = \text{Tr} \left( (\bm{Y} - \bm{PQ}^H \bm{X})(\bm{Y} -
      \bm{PQ}^H \bm{X})^H \right)
  \end{aligned}
\]
for the sake of notational simplicity, the solution
\( \left(\bm{P}^*, \bm{Q}^* \right) \) of Eq.~\eqref{eq: appendix --
  dmd norm minimization} is given by the stationary points of
\( \mathcal{J}(\bm{P}, \bm{Q}) \), i.e.\
\[
  \begin{aligned}
    \frac{\partial \mathcal{J}}{\partial \bm{P}} & = -2 \bm{YX}^H \bm{Q} + 2 \bm{PQ}^H \bm{XX}^H \bm{Q} & = 0\\
    \frac{\partial \mathcal{J}}{\partial \bm{Q}} & = -2 \bm{XY}^H \bm{P} + 2 \bm{XX}^H \bm{QP}^H \bm{P} & = 0
  \end{aligned}
\]
Introducing the orthogonality constraint on \( \bm{P} \) in the second
equation above yields the following expression for \( \bm{Q} \)
\[
  \bm{Q}^H = \bm{P}^H \bm{C}_{yx} \bm{C}_{xx}^{-1},
\]
where \( \bm{C}_{yx} = \bm{YX}^H \) and \( \bm{C}_{xx} = \bm{XX}^H \)
are the empirical variance-covariance matrices introduced in
\textsection~\ref{subsec: numerics -- dmd}.  Inserting this expression
for \( \bm{Q} \) in Eq.~\eqref{eq: appendix -- dmd norm minimization}
yields
\begin{equation}
  \begin{aligned}
    \minimize_{\bm{P}} & \| \bm{Y} - \bm{PP}^H \bm{C}_{yx} \bm{C}_{xx}^{-1} \bm{X} \|_F^2 \\
    \subjectto & \text{rank } \bm{P} = r \\
    & \bm{P}^H \bm{P} = \bm{I},
  \end{aligned}
  \label{eq: appendix -- dmd norm minimization bis}
\end{equation}
From this point, it is straightforward to show that the above problem
can be rewritten as
\begin{equation}
  \begin{aligned}
    \minimize_{\bm{P}} & \text{Tr}\left( \bm{C}_{yy} \right) - 2 \text{Tr}\left( \bm{P}^H \bm{C}_{yx} \bm{C}_{xx}^{-1} \bm{C}_{xy} \bm{P} \right) \\
    \subjectto & \text{rank } \bm{P} = r \\
    & \bm{P}^H \bm{P} = \bm{I}.
  \end{aligned}
  \label{eq: appendix -- dmd norm minimization bis}
\end{equation}
\( \text{Tr}\left( \bm{C}_{yy} \right) \) being constant, we thus
established the equivalence between the norm minimization
problem~\eqref{eq: appendix -- dmd norm minimization} and the
following trace maximization problem
\begin{equation}
  \begin{aligned}
    \maximize_{\bm{P}} & \text{Tr}\left( \bm{P}^H \bm{C}_{yx} \bm{C}_{xx}^{-1} \bm{C}_{xy} \bm{P} \right) \\
    \subjectto & \text{rank } \bm{P} = r \\
    & \bm{P}^H \bm{P} = \bm{I}.
  \end{aligned}
  \label{eq: appendix -- dmd trace maximization}
\end{equation}
Finally, recalling that the trace of a matrix is the sum of its
eigenvalues, basic linear algebra is sufficient to prove that the
solution of the above optimizations problem is given by the first
\(r \) leading eigenvectors of the symmetric positive definite matrix
\( \bm{C}_{yx} \bm{C}_{xx}^{-1} \bm{C}_{xy} \).  Once \( \bm{P} \) has
been computed, the matrix \( \bm{Q} \) is simply given by
\( \bm{Q}^H = \bm{P}^H \bm{C}_{yx} \bm{C}_{xx}^{-1} \) as stated
previously.

Interestingly, it can be noted that if \( \bm{X} \) is a full rank
\( m \times m \) matrix (i.e.\ as many linearly independent samples as
degrees of freedom), then
\[
  \begin{aligned}
    \bm{C}_{yx} \bm{C}_{xx}^{-1} \bm{C}_{xy} & = \bm{Y} \bm{X}^H \left( \bm{XX}^H \right)^{-1} \bm{XY}^H \\
    & = \bm{YY}^H \\
    & = \bm{C}_{yy}.
  \end{aligned}
\]
In this particular case (hardly encountered in fluid dynamics due to
the memory footprint of storing \( m \) snapshots of a
\( m \)-dimensional vector), the columns of \( \bm{P} \) are simply
the first \( r \) leading eigenvectors of the variance-covariance
matrix \( \bm{C}_{yy} \), i.e.\ the POD modes of the output matrix
\( \bm{Y} \).  Alternatively, if \( \bm{X} \) is a skinny
\(m \times n\) matrix (with \( m > n \)) or a rank-deficient
\( m \times m \) matrix, one can introduce its economy-size singular
value decomposition
\[
  \bm{X} = \bm{U}_{\bm{X}} \boldsymbol{\Upsigma}_{\bm{X}}
  \bm{V}^H_{\bm{X}}.
\]
The columns of \( \bm{P} \) are then given by the first \( r \) left
singular vectors of the matrix \( \bm{YV}_{\bm{X}}\bm{V}_{\bm{X}}^H \)
where \( \bm{V}_{\bm{X}} \bm{V}_{\bm{X}}^H \) is the projector onto
the rowspan of \( \bm{X} \).

\subsection{Computing the DMD modes and eigenvalues from the low-rank
  factorization of \( \bm{A} \)}

It must be noted that the columns of \( \bm{P} \) are not the
so-called DMD modes \( \boldsymbol{\Uppsi} \) although they span the
same subspace.  The DMD modes can nonetheless be easily computed from
the low-rank factorization \( \bm{A} = \bm{PQ}^H \) of the DMD
operator.  The i\textsuperscript{th} DMD mode is solution to the
eigenvalue problem
\[
  \boldsymbol{\Uppsi}_i \lambda_i = \bm{A} \boldsymbol{\Uppsi}_i.
\]
Since \( \boldsymbol{\Uppsi}_i \) needs to be in the columnspan of
\( \bm{A} \), it can be expressed as a linear combination of the
columns of \( \bm{P} \), i.e.\
\[
  \boldsymbol{\Uppsi}_i = \bm{P} \bm{b}_i
\]
where \( \bm{b}_i \) is a small \( r \)-dimensional vector.
Introducing this expression into the DMD eigenvalue problem yields
after some simplifications
\[
  \bm{b}_i \lambda_i = \bm{Q}^H \bm{P} \bm{b}_i.
\]
Although \( \bm{A} \) is a high-dimensional \( m \times m \) matrix,
\( \bm{Q}^H \bm{P} \) is a small \( r \times r \) matrix whose
eigenvectors \( \bm{b}_i \) and eigenvalues \( \lambda_i \) can easily
be computed using direct solvers thus enabling the computations of the
eigenvalues and eigenvectors of \( \bm{A} \) at a reduced cost.

Similarly, the \emph{adjoint} DMD modes \( \boldsymbol{\Upphi}_i \)
are solution to the adjoint DMD eigenvalue problem
\[
  \boldsymbol{\Upphi}_i \bar{\lambda}_i = \bm{A}^H
  \boldsymbol{\Upphi}_i,
\]
where the overbar denotes the complex conjugate.  These now live in
the rowspan of \( \bm{A} \) and can thus be expressed as
\[
  \boldsymbol{\Upphi}_i = \bm{Qc}_i
\]
where once again \( \bm{c}_i \) is a small \( r \)-dimensional vector.
The corresponding low-dimensional eigenvalue problem then reads
\[
  \bm{c}_i \bar{\lambda}_i = \bm{P}^H \bm{Q} \bm{c}_i
\]
with \( \bm{P}^H \bm{Q} \) a small \( r \times r \) matrix.  It must
be noted however that, from a practical point of view, converging the
adjoint DMD modes \( \boldsymbol{\Upphi}_i \) may require
significantly more data than needed to the converge the direct DMD
modes \( \boldsymbol{\Uppsi}_i \).
 \section{Broomhead-King embedding for attractor
  reconstruction}\label{appendix: broomhead-king}

Attractor reconstruction from a single time-series has been
instrumental in highlighting the connection between the dynamics of
the thermosyphon and that of the canonical Lorenz system (see
figure~\ref{fig: flow configuration -- flow rate characterization} and
figure~\ref{fig: flow configuration -- lorenz system}).  The aim of
this section is to briefly introduce inexperienced readers to the
Broomhead-King attractor reconstruction technique used herein.  For
more details, please refer to the original
paper~\cite{physicaD:broomhead:1986}.  For the sake of simplicity and
reproducibility, we will apply this attractor reconstruction technique
using the Lorenz system
\[
  \begin{aligned}
    \dot{x} & = \sigma \left( y - x \right) \\
    \dot{y} & = x \left( \rho - z \right) - y \\
    \dot{z} & = xy - \beta z
  \end{aligned}
\]
with
\( \left( \sigma, \rho, \beta \right) = \left( 10, 28, \nicefrac{8}{3}
\right) \) so that it exhibits chaotic dynamics.  The properties of
this dynamical system have been discussed in \textsection~\ref{sec:
  lorenz}.  Figure~\ref{fig: appendix -- lorenz attractor} depicts its
well-known strange attractor along with representative time-series of
\( x(t) \), \( y(t) \) and \( z(t) \).  In the rest of this appendix,
we will work exclusively with the time-series of \( x(t) \).  The
whole dataset consists of the evolution of \( x(t) \) from \(t = 0\)
to \(t = 400\) with a sampling period \( \Delta t = 0.001 \) resulting
in 400 000 samples.

Although the Takens embedding theorem~\cite{takens:1981} provides
theoretical guarantees to reconstruct a multidimensional attractor
from a single time-series, this approach might break when applied to
real data that may be contaminated by measurement noise or it might be
quite sensitive on the sampling period used to acquire the signal.
Additionally, once the lag \( \tau \) has been determined, the number
of time-lagged versions of \( x(t) \) needed to reconstruct the
attractor needs to be determined using techniques based nearest
neighbors~\cite{pra:kennel:1992,physicaD:cao:1997,jcs:krakovska:2015}
which might provide conflicting estimates of the attractor's dimension
and different reconstructions.  As to overcome these limitations,
Broomhead \& King~\cite{physicaD:broomhead:1986} have proposed instead
to reconstruct the attractor based on the singular value decomposition
of the Hankel matrix constructed from time-lagged version of
\( x(t) \)
\[
  \bm{H} = \frac{1}{\sqrt{N}} \begin{bmatrix}
    x_1 & x_2 & x_3 & \cdots & x_d \\
    x_2 & x_3 & x_4 & \cdots & x_{d+1} \\
    \vdots & \vdots & \vdots & \cdots & \vdots \\
    x_{N-d+1} & x_{N-d+2} & x_{N-d+3} & \cdots & x_{N} \\
  \end{bmatrix}
\]
where \( x_k = x(t_k) \), \( N \) is the total length of the
time-series (i.e.\ 400 000 in our case) and \( d \) is the maximum
number of time-lagged considered.  The choice of \( d \) will be
discussed later on.  Decomposing this Hankel matrix as
\( \bm{H} = \bm{U} \boldsymbol{\Upsigma} \bm{V}^H \), the dimension of
the attractor can then be inferred from the distribution of the
singular values \( \sigma_i \).  The columns of \( \bm{U} \) then
provide an orthonormal basis for the embedding space while the columns
of \( \bm{V} \) describe the evolution of the system within this
embedding space, thus enabling the reconstruction we aimed for.

The choice of the number of lags \( d \) needed to construct the
Hankel matrix is of crucial importance as it will determine the window
length \( \tau_w = d \Delta t\) used to embed the dynamics which, in
turn, will determine the structure of the singular value spectrum.  In
practice, the window length \( \tau_w \) needs to be large enough to
average out the possible noise contaminating the data.  Note however
that, in the limit \( \tau_w \to \infty \), the singular value
decomposition of the Hankel matrix converges to the discrete Fourier
transform of \( x(t) \).  Chaotic systems being characterized by a
continuous spectrum, an infinite number of singular components would
thus be needed to reconstruct the signal, grossly overestimating the
dimension of the attractor.  In their paper, Broomhead \&
King~\cite{physicaD:broomhead:1986} recommend to use a window length
\( \tau_w \) of approximately one tenth of the period of oscillation
about the unstable saddle-foci.  In this work, this window length has
been chosen as one twentieth of this oscillating period, even though
choosing one tenth does not change qualitatively (nor quantitatively)
the results presented.  Note that the same rationale has been used
when applying this technique to the data from the thermosyphon.  For
our choice of parameters, figure~\ref{fig: appendix -- broomhead king}
provides the temporal evolution of the system within the embedding
space and the corresponding reconstruction of the attractor.  The
theoretical properties of such time-delay embedding of attractors have
been recently studied by Kamb \emph{et al.}~\cite{arxiv:kamb:2018}.

  \begin{figure}
    \centering \includegraphics[width=\textwidth]{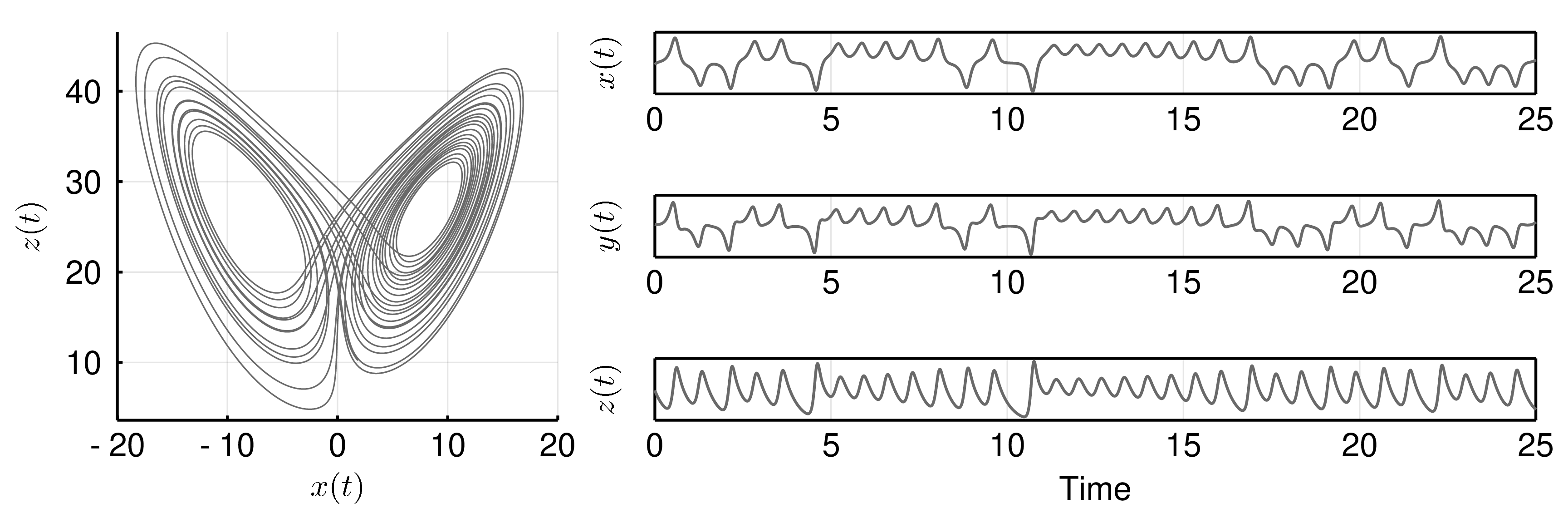}
    \caption{ Projection of the Lorenz attractor on the \((x, z)\)
      plane (left) and corresponding time-series of \( x(t) \),
      \(y(t)\) and \(z(t)\) (right).  The parameters of the system are
      chosen as
      \( \left( \sigma, \rho, \beta \right) = \left( 10, 28,
        \nicefrac{8}{3} \right) \).  }\label{fig: appendix -- lorenz
      attractor}
  \end{figure}

  \begin{figure}
    \centering \includegraphics[width=\textwidth]{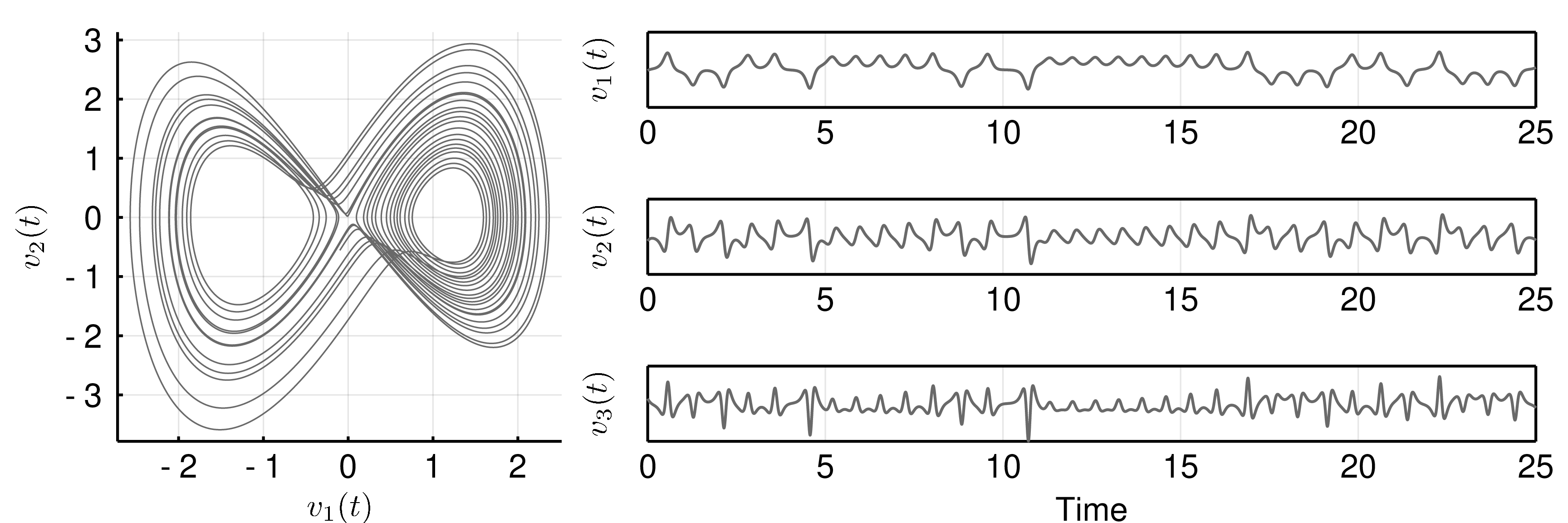}
    \caption{ Projection of the Lorenz attractor reconstructed using
      the Broomhead-King technique on the \((v_1, v_2)\) plane (left)
      and corresponding time-series of \( v_1(t) \), \(v_2(t)\) and
      \(v_3(t)\) (right).  A window length \( \tau_w = 0.032 \),
      corresponding to roughly one twentieth of the oscillating period
      around the unstable saddle-foci, has been chosen.  }\label{fig:
      appendix -- broomhead king}
  \end{figure}


\bibliographystyle{spphys} 
\bibliography{bibliography} 

\begin{thebibliography}{10}
\providecommand{\url}[1]{{#1}}
\providecommand{\urlprefix}{URL }
\expandafter\ifx\csname urlstyle\endcsname\relax
  \providecommand{\doi}[1]{DOI \discretionary{}{}{}#1}\else
  \providecommand{\doi}{DOI \discretionary{}{}{}\begingroup
  \urlstyle{rm}\Url}\fi

\bibitem{book:holmes:1996}
P.~Holmes, J.L. Lumley, G.~Berkooz, \emph{Turbulence, Coherent Structures,
  Dynamical Systems and Symmetry} (Cambridge University Press, 1996).
\newblock \doi{10.1017/CBO9780511622700}

\bibitem{amr:fabiane:2014}
N.~Fabbiane, O.~Semeraro, S.~Bagheri, D.S. Henningson, Appl. Mech. Rev.
  (2014).
\newblock \doi{10.1115/1.4027483}

\bibitem{amr:brunton:2015}
S.L. Brunton, B.R. Noack, Appl. Mech. Rev. \textbf{67}(5), 050801 (2015).
\newblock \doi{10.1115/1.4031175}

\bibitem{amr:sipp:2016}
D.~Sipp, P.J. Schmid, Appl. Mech. Rev. \textbf{68}(2), 020801 (2016).
\newblock \doi{10.1115/1.4033345}

\bibitem{arfm:rowley:2017}
C.W. Rowley, S.T.M. Dawson, Annu. Rev. Fluid Mech. \textbf{49}(1), 387 (2017).
\newblock \doi{10.1146/annurev-fluid-010816-060042}

\bibitem{aiaa:taira:2017}
K.~Taira, S.L. Brunton, S.T.M. Dawson, C.W. Rowley, T.~Colonius, B.J. McKeon,
  O.T. Schmidt, S.~Gordeyev, V.~Theofilis, L.S. Ukeiley, {AIAA} J.
  \textbf{55}(12), 4013 (2017).
\newblock \doi{10.2514/1.J056060}

\bibitem{jas:lorenz:1963}
E.N. Lorenz, J. Atmos. Sci. \textbf{20}(2), 130 (1963).
\newblock \doi{https://doi.org/10.1175/1520-0469(1963)020<0130:DNF>2.0.CO;2}

\bibitem{qam:sirovich:1987}
L.~{Sirovich}, Q. Appl. Math. \textbf{45}, 561 (1987).
\newblock \doi{10.1090/qam/910462}

\bibitem{amr:berkooz:1993}
G.~Berkooz, P.~Holmes, J.L. Lumley, Annu. Rev. Fluid Mech. \textbf{25}(1), 539
  (1993).
\newblock \doi{10.1146/annurev.fl.25.010193.002543}

\bibitem{jfm:noack:2003}
B.R. Noack, K.~Afanasiev, M.~Morzy{\'{n}}ski, G.~Tadmor, F.~Thiele, J. Fluid
  Mech. \textbf{497}, 335 (2003).
\newblock \doi{10.1017/s0022112003006694}

\bibitem{jgcd:juang:1985}
J.N. Juang, R.S. Pappa, J. Guidance \textbf{8}(5), 620 (1985).
\newblock \doi{10.2514/3.20031}

\bibitem{jfm:rowley:2009}
C.W. Rowley, I.~Mezi{\'{c}}, S.~Bagheri, P.~Schlatter, J. Fluid Mech.
  \textbf{641}, 115 (2009).
\newblock \doi{10.1017/S0022112009992059}

\bibitem{jfm:schmid:2010}
P.J. Schmid, J. Fluid Mech. \textbf{656}, 5 (2010).
\newblock \doi{10.1017/S0022112010001217}

\bibitem{jcd:tu:2014}
J.H. Tu, C.W. Rowley, D.M. Luchtenburg, S.L. Brunton, J.N. Kutz, J. Comp. Dyn.
  \textbf{1}(2), 391 (2014).
\newblock \doi{10.3934/jcd.2014.1.391}

\bibitem{book:billings:2013}
S.A. Billings, \emph{{Nonlinear System Identification: NARMAX Methods in the
  Time, Frequency, and Spatio-Temporal Domains}} (Paperbackshop UK Import,
  2013).
\newblock
  \urlprefix\url{https://www.ebook.de/de/product/20764167/stephen_a_billings_nonlinear_system_identification_narmax_methods_in_the_time_frequency_and_spatio_temporal_domains.html}

\bibitem{pnas:bongard:2007}
J.~Bongard, H.~Lipson, Proc. Natl. Acad. Sci. U.S.A. \textbf{104}(24), 9943
  (2007).
\newblock \doi{10.1073/pnas.0609476104}

\bibitem{science:schmidt:2009}
M.~Schmidt, H.~Lipson, Science \textbf{324}(5923), 81 (2009).
\newblock \doi{10.1126/science.1165893}

\bibitem{pnas:brunton:2016}
S.L. Brunton, J.L. Proctor, J.N. Kutz, Proc. Natl. Acad. Sci. U.S.A.
  \textbf{113}(15), 3932 (2016).
\newblock \doi{10.1073/pnas.1517384113}

\bibitem{takens:1981}
F.~Takens, in \emph{Lecture Notes in Mathematics} (Springer Berlin Heidelberg,
  1981), pp. 366--381.
\newblock \doi{10.1007/BFb0091924}

\bibitem{physicaD:broomhead:1986}
D.S. Broomhead, G.P. King, Physica D \textbf{20}(2-3), 217 (1986).
\newblock \doi{10.1016/0167-2789(86)90031-x}

\bibitem{physd:yorke:1987}
J.A. Yorke, E.D. Yorke, J.~Mallet-Paret, Physica D \textbf{24}(1-3), 279 (1987)

\bibitem{jfm:ehrhard:1990}
P.~Ehrhard, U.~M{\"u}ller, J. Fluid Mech. \textbf{217}, 487 (1990)

\bibitem{ijhmt:louisos:2015}
W.F. Louisos, D.L. Hitt, C.M. Danforth, Int. J. Heat Mass Transfer \textbf{88},
  492 (2015)

\bibitem{physicaA:anishchenko:2003}
V.S. Anishchenko, T.E. Vadivasova, G.A. Okrokvertskhov, G.I. Strelkova, Physica
  A \textbf{325}(1-2), 199 (2003)

\bibitem{jas:saltzman:1962}
B.~Saltzman, J. Atmos. Sci. \textbf{19}(4), 329 (1962)

\bibitem{pof:jovanovic:2014}
M.R. Jovanovi{\'c}, P.J. Schmid, J.W. Nichols, Phys. Fluids \textbf{26}(2),
  024103 (2014)

\bibitem{jns:williams:2015}
M.O. Williams, I.G. Kevrekidis, C.W. Rowley, J. Nonlinear Sci. \textbf{25}(6),
  1307 (2015).
\newblock \doi{10.1007/s00332-015-9258-5}

\bibitem{book:kutz:2016}
J.N. Kutz, S.L. Brunton, B.W. Brunton, J.L. Proctor, \emph{Dynamic Mode
  Decomposition: Data-Driven Modeling of Complex Systems} (SIAM-Society for
  Industrial and Applied Mathematics, 2016).
\newblock
  \urlprefix\url{https://www.amazon.com/Dynamic-Mode-Decomposition-Data-Driven-Modeling/dp/1611974496?SubscriptionId=AKIAIOBINVZYXZQZ2U3A&tag=chimbori05-20&linkCode=xm2&camp=2025&creative=165953&creativeASIN=1611974496}

\bibitem{expfluids:dawson:2016}
S.T.M. Dawson, M.S. Hemati, M.O. Williams, C.W. Rowley, Exp. Fluids
  \textbf{57}(3) (2016).
\newblock \doi{10.1007/s00348-016-2127-7}

\bibitem{jfm:noack:2016}
B.R. Noack, W.~Stankiewicz, M.~Morzy{\'{n}}ski, P.J. Schmid, J. Fluid Mech.
  \textbf{809}, 843 (2016).
\newblock \doi{10.1017/jfm.2016.678}

\bibitem{siam:leclainche:2017}
S.~Le~Clainche, J.M. Vega, {SIAM} Journal on Applied Dynamical Systems
  \textbf{16}(2), 882 (2017).
\newblock \doi{10.1137/15m1054924}

\bibitem{arxiv:hirsh:2019}
S.M. Hirsh, K.D. Harris, J.N. Kutz, B.W. Brunton,   (2019)

\bibitem{pnas:koopman:1932}
B.O. Koopman, Proc. Natl. Acad. Sci. U.S.A. \textbf{17}(5), 315 (1931)

\bibitem{jma:izenman:1975}
A.J. Izenman, Journal of multivariate analysis \textbf{5}(2), 248 (1975)

\bibitem{ieee:delatorre:2012}
F.~De~la Torre, IEEE Transactions on Pattern Analysis and Machine Intelligence
  \textbf{34}(6), 1041 (2012)

\bibitem{arxiv:heas:2016}
P.~H{\'e}as, C.~Herzet, arXiv e-print:1610.02962  (2016)

\bibitem{arxiv:tegmark:2019}
M.~Tegmark, arXiv e-print:1902.03364  (2019)

\bibitem{jfm:loiseau:2018a}
J.C. Loiseau, S.L. Brunton, J. Fluid Mech. \textbf{838}, 42 (2018).
\newblock \doi{10.1017/jfm.2017.823}

\bibitem{Tibshirani1996lasso}
R.~Tibshirani, Journal of the Royal Statistical Society. Series B
  (Methodological) \textbf{73}, 267 (1996).
\newblock \doi{10.1111/j.1467-9868.2011.00771.x}

\bibitem{ieee:mangan:2016}
N.M. Mangan, S.L. Brunton, J.L. Proctor, J.N. Kutz, IEEE Trans. Mol. Biol.
  Multi-Scale Commun. \textbf{2}(1), 52 (2016).
\newblock \doi{10.1109/tmbmc.2016.2633265}

\bibitem{ifac:brunton:2016}
S.L. Brunton, J.L. Proctor, J.N. Kutz, {IFAC}-{PapersOnLine} \textbf{49}(18),
  710 (2016).
\newblock \doi{10.1016/j.ifacol.2016.10.249}

\bibitem{nature:brunton:2017}
S.L. Brunton, B.W. Brunton, J.L. Proctor, E.~Kaiser, J.N. Kutz, Nat. Commun.
  \textbf{8}(1) (2017).
\newblock \doi{10.1038/s41467-017-00030-8}

\bibitem{sciadv:rudy:2017}
S.H. Rudy, S.L. Brunton, J.L. Proctor, J.N. Kutz, Sci. Adv. \textbf{3}(4),
  e1602614 (2017).
\newblock \doi{10.1126/sciadv.1602614}

\bibitem{arxiv:kaiser:2017}
E.~Kaiser, J.N. Kutz, S.L. Brunton, ArXiv e-prints  (2017)

\bibitem{jfm:loiseau:2018b}
J.C. Loiseau, B.R. Noack, S.L. Brunton, J. Fluid Mech. \textbf{844}, 459
  (2018).
\newblock \doi{10.1017/jfm.2018.147}

\bibitem{pof:john:2012}
J.~John Soundar~J., J.M. Chomaz, P.~Huerre, Phys. Fluids \textbf{24}(4), 044103
  (2012)

\bibitem{pof:podvin:2012}
B.~Podvin, A.~Sergent, Phys. Fluids \textbf{24}(10), 105106 (2012)

\bibitem{jfm:podvin:2019}
A.~Castillo-Castellanos, A.~Sergent, B.~Podvin, M.~Rossi, J. Fluid Mech.
  \textbf{877}, 922–954 (2019).
\newblock \doi{10.1017/jfm.2019.598}

\bibitem{siam:proctor:2016}
J.L. Proctor, S.L. Brunton, J.N. Kutz, SIAM J. Appl. Dyn. Syst. \textbf{15}(1),
  142 (2016)

\bibitem{ieee:manohar:2018}
K.~Manohar, B.W. Brunton, J.N. Kutz, S.L. Brunton, IEEE Control Syst. Mag.
  \textbf{38}(3), 63 (2018).
\newblock \doi{10.1109/MCS.2018.2810460}

\bibitem{ieee:clark:2019}
E.~{Clark}, T.~{Askham}, S.L. {Brunton}, J.~{Nathan Kutz}, IEEE Sensors Journal
  \textbf{19}(7), 2642 (2019).
\newblock \doi{10.1109/JSEN.2018.2887044}

\bibitem{pra:kennel:1992}
M.B. Kennel, R.~Brown, H.D.I. Abarbanel, Phys. Rev. A \textbf{45}, 3403 (1992).
\newblock \doi{10.1103/PhysRevA.45.3403}

\bibitem{physicaD:cao:1997}
L.~Cao, Physica D: Nonlinear Phenomena \textbf{110}(1), 43  (1997).
\newblock \doi{10.1016/S0167-2789(97)00118-8}.
\newblock
  \urlprefix\url{http://www.sciencedirect.com/science/article/pii/S0167278997001188}

\bibitem{jcs:krakovska:2015}
A.~Krakovsk{\'a}, K.~Mezeiov{\'a}, H.~Bud{\'a}{\v{c}}ov{\'a}, Journal of
  Complex Systems \textbf{2015} (2015)

\bibitem{arxiv:kamb:2018}
M.~Kamb, E.~Kaiser, S.L. Brunton, J.N. Kutz, arXiv preprint arXiv:1810.01479
  (2018)

\end{thebibliography}

\end{document}